\documentclass[rmp,aps,twocolumn,nofootinbib,floatfix]{revtex4} 
\usepackage{epsf}
\usepackage{graphics}
\usepackage{epsfig}
\usepackage{dcolumn}
\usepackage{bm}
\usepackage[dvips]{color}

\def\be{\begin{equation}}
\def\ee{\end{equation}}
\def\ba{\begin{eqnarray}}
\def\ea{\end{eqnarray}}

\begin{document}

\title{Transport in strongly correlated two dimensional electron fluids}

\author{B. Spivak}
\affiliation{Department of Physics, University of Washington, Seattle, Washington 98195, USA}

\author{S.~V. Kravchenko}
\affiliation{Physics Department, Northeastern University, Boston, Massachusetts 02115, USA}

\author{S.~A. Kivelson}
\affiliation{Department of Physics, Stanford University, Stanford, California 94305, USA}

\author{X.~P.~A. Gao}
\affiliation{Department of Physics, Case Western Reserve University, Cleveland, Ohio 44106, USA}

\begin{abstract}
We present an overview of the measured transport properties of the two dimensional electron fluids in high mobility semiconductor devices with low electron densities, and of some of the theories that have been proposed to account for them.  Many features of the observations are not easily reconciled with a description based on the well  understood physics of weakly interacting quasiparticles in a disordered medium.  Rather, they reflect new physics associated with strong correlation effects, which warrant further study.
 \end{abstract}


\maketitle

\tableofcontents\vspace{1 cm}

As low density  two dimensional (2D) electronic systems with
increasingly high mobility have become available, there has
accumulated experimental evidence of a set of low temperature
phenomena which cannot be easily
understood on the basis of
traditional (Fermi liquid based) metal physics.\footnote{For experimental results on various 2D structures, see \textcite{zavaritskaya87},
\textcite{diorio90}, \textcite{shashkin93}, \textcite{krav}, \textcite{kravchenko95a}, \textcite{kravchenko96}, \textcite{popovic97}, \textcite{kravHpar}, \textcite{pudalov97}, \textcite{coleridge97}, \textcite{pudalov}, \textcite{pudHpar}, \textcite{papadakis98}, \textcite{hanein98}, \textcite{simmons98}, \textcite{dultz98}, \textcite{hanein98a}, \textcite{kravchenko98}, \textcite{sar}, \textcite{yoon99}, \textcite{okamoto}, \textcite{ensslin}, \textcite{feng99}, \textcite{mertes99}, \textcite{mills99}, \textcite{hanein99}, \textcite{simmons00}, \textcite{kravchenko00}, \textcite{yacoby}, \textcite{Shahar}, \textcite{Jiang}, \textcite{kravchenko00a}, \textcite{AbrKravSar}, \textcite{feng01}, \textcite{shashkin01a}, \textcite{vitkalov01a}, \textcite{fletcher}, \textcite{gao01}, \textcite{gao02}, \textcite{PudalovSpin}, \textcite{shashkin02a}, \textcite{shashkin02b}, \textcite{depoortere02}, \textcite{popovic02}, \textcite{noh03}, \textcite{StormerZhuSpin}, \textcite{eisenstein}, \textcite{Pillarisetty}, \textcite{shashkin03}, \textcite{GershenzonPudalov}, \textcite{GaoLanl}, \textcite{prus}, \textcite{okamoto03}, \textcite{ShayeganSpin}, \textcite{kravchenko04}, \textcite{lai05}, \textcite{Gao2005}, \textcite{shashkin05}, \textcite{Sarma1}, \textcite{VitSat}, \textcite{GaoHpar05}, \textcite{shashkin06}, \textcite{lai07}, \textcite{anissimova07}, \textcite{popovic07}, \textcite{TsuiSpin}, and \textcite{Kane09}}
More precisely, these are systems with a
large ratio between the typical potential and the kinetic
energies, $r_{\text s} \equiv 1/\sqrt{\pi n(a_{\text B})^2}$, where $n$ is
the areal density of electrons and $a_{\text B}^*=\hbar^2 \epsilon/m^*e^2$
is the effective Bohr radius. A generic feature of electronic
systems is a low temperature evolution as a function of $n$ from a
conducting ``Fermi liquid'' (FL) state at large $n$ to an
insulating ``Wigner crystalline'' (WC) state at low.
To obtain as clear a perspective as possible on this physics, we
focus  on 2D electron (and hole) systems in which $r_{\text s}$ is large, so
interaction effects manifestly cannot be treated perturbatively, but in which the system  remains a {\it fluid},
hence thermodynamically distinct from a WC.

It should be stressed that while the experimental data  exhibits striking
features which have been confirmed by multiple groups on various materials,
there is disagreement  concerning the correct theoretical perspective from
which to view these experiments.  Of the unsettled issues, the most vexed is the
proper theoretical interpretation of the experimentally observed metal-insulator
transition.
There is also much that remains to be understood concerning the observed
transport anomalies in samples which are far  from the
critical region.
Here we summarize the salient experimental facts,
stressing the anomalous character of the results obtained in conditions  both
close to and far from
the metal-insulator transition.  We also summarize some of the major theoretical
approaches, including a discussion of both their successes and
shortcomings.  The primary purpose of this Colloquium is to stimulate
further experimental work on this subject,
particularly work aimed at
unraveling the
non-critical behavior of the metallic and insulating states that occur in
strongly correlated two dimensional electron fluids far from the metal-insulator
transition.

\section{Experimental signatures of Non-Fermi liquid behavior}
\label{experiment}

In this section, we summarize some of the experimental results on
large $r_{\text s}$ 2D electron and hole gases (2DEG's and 2DHG's) in
highly conducting semiconductor heterostructures that we contend
are incompatible with a Fermi liquid based theory and with the
single particle  theory of weak localization.  We will
particularly focus on experiments on the following systems:  n-Si
MOSFETs,\footnote{Transport and thermodynamic properties of dilute silicon MOSFETs have been extensively by \textcite{zavaritskaya87},\textcite{diorio90}, \textcite{shashkin93}, \textcite{krav}, \textcite{kravchenko95a}, \textcite{kravchenko96}, \textcite{popovic97}, \textcite{kravHpar}, \textcite{pudalov97}, \textcite{pudalov}, \textcite{pudHpar}, \textcite{kravchenko98}, \textcite{okamoto}, \textcite{feng99}, \textcite{mertes99}, \textcite{kravchenko00}, \textcite{kravchenko00a}, \textcite{feng01}, \textcite{shashkin01a}, \textcite{vitkalov01a}, \textcite{fletcher}, \textcite{PudalovSpin}, \textcite{shashkin02a}, \textcite{popovic02}, \textcite{shashkin02b}, \textcite{shashkin03}, \textcite{GershenzonPudalov}, \textcite{prus}, \textcite{VitSat}, \textcite{shashkin06}, \textcite{anissimova07}, \textcite{popovic07}, and \textcite{Kane09}}
p-GaAs hetero-junctions and quantum-wells,\footnote{For experiments performed on p-GaAs structures, see
\textcite{hanein98}, \textcite{simmons98}, \textcite{dultz98}, \textcite{mills99}, \textcite{yoon99}, \textcite{hanein99}, \textcite{simmons00}, \textcite{yacoby}, \textcite{Shahar}, \textcite{Jiang}, \textcite{gao01}, \textcite{gao02}, \textcite{noh03}, \textcite{GaoLanl}, \textcite{Pillarisetty}, \textcite{Gao2005}, and \textcite{GaoHpar05}}
n-GaAs hetero-junctions,\footnote{Dilute electron gases in n-GaAs hetero-junctions have been studied by
\textcite{hanein98a}, \textcite{StormerZhuSpin}, \textcite{eisenstein}, and \textcite{Sarma1}} p-SiGe quantum
wells \cite{coleridge97,ensslin}, AlAs quantum wells \cite{papadakis98,depoortere02,ShayeganSpin}, and n-SiGe
structures (\textcite{okamoto03}, \textcite{lai05}, \textcite{lai07}, \textcite{TsuiSpin}). To orient the
reader,  the best current theoretical estimates
(\textcite{cip}; \textcite{attaccalite}) of the critical value of $r_{\text s}$ at which the
energies of a uniform FL and WC are equal is $r_{\text s}^\star \approx
38$, while for the devices we have in mind, $r_{\text s}\sim 5-20$ in
the case of the Si MOSFETs and $\sim 10-40$  for the p-GaAs
devices.

One of the key points to notice as we discuss the data below is the great similarities in the structure of the data from the different devices.  This implies that the observed anomalies represent robust ``universal" behaviors of the 2DEG, which are largely independent of details.  This is correct in spite of the fact that there are significant differences between the electronic structures of the various devices. In p-GaAs and n-GaAs heterostructures, the electrons or holes occupy a single band with an isotropic effective mass, while in n-Si MOSFETs, there are two degenerate valleys and correspondingly a non-trivial structure to the effective mass tensor.   In the heterostructures, the interactions between electrons at large separation are Coulombic, while in MOSFETs, the interaction between electrons is dipolar at distances large compared to the distance to the metal gate.  In p-GaAs and p-SiGe, the spin-orbit coupling may be significant, while in n-GaAs and Si MOSFETs it is clearly insignificant.
 In Si MOSFETs the disorder potential is believed to have  short-ranged correlations, while in most of the other 2D systems considered here, it is believed to be long-range correlated ({\it i.e.} due, primarily, to distant charged impurities).
\begin{figure*}[h]
         \large(a)\includegraphics[width=0.32\textwidth]{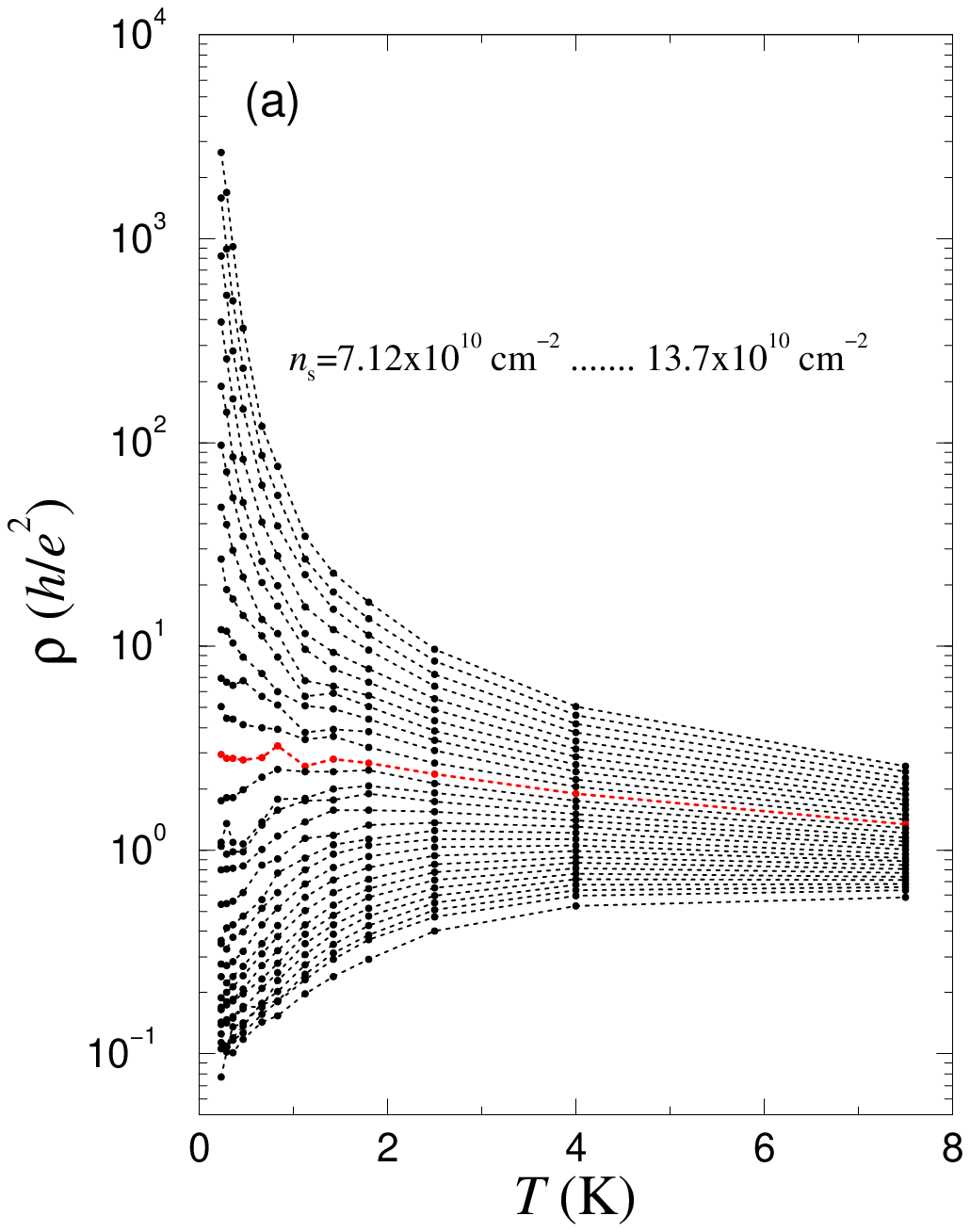}
        \large(b)\includegraphics[viewport=0 -25 500 582,width=0.34\textwidth,clip]{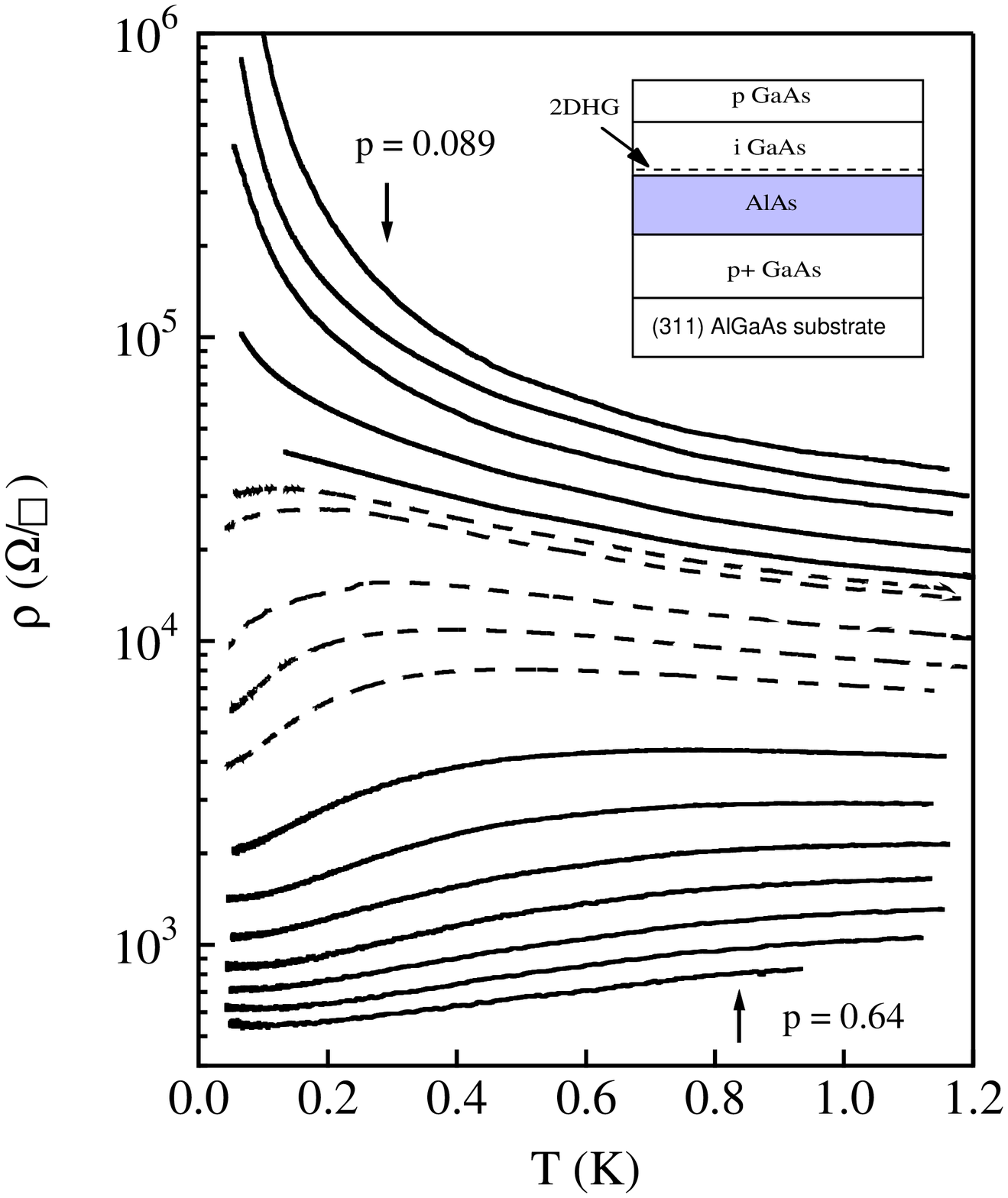}
         \large(c)\includegraphics[width=0.32\textwidth]{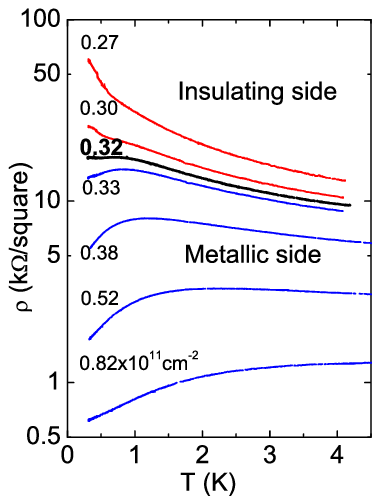}
        \large(d)\includegraphics[viewport=0 -10 227 252,width=0.32\textwidth,clip]{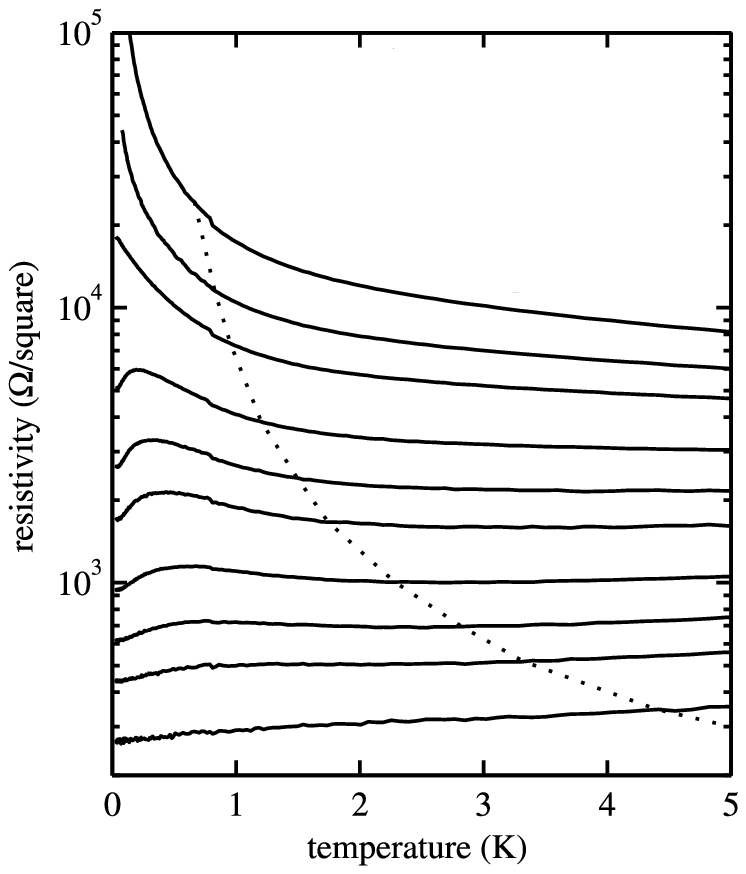}\vspace*{3mm}
         \large(e)\includegraphics[width=0.35\textwidth]{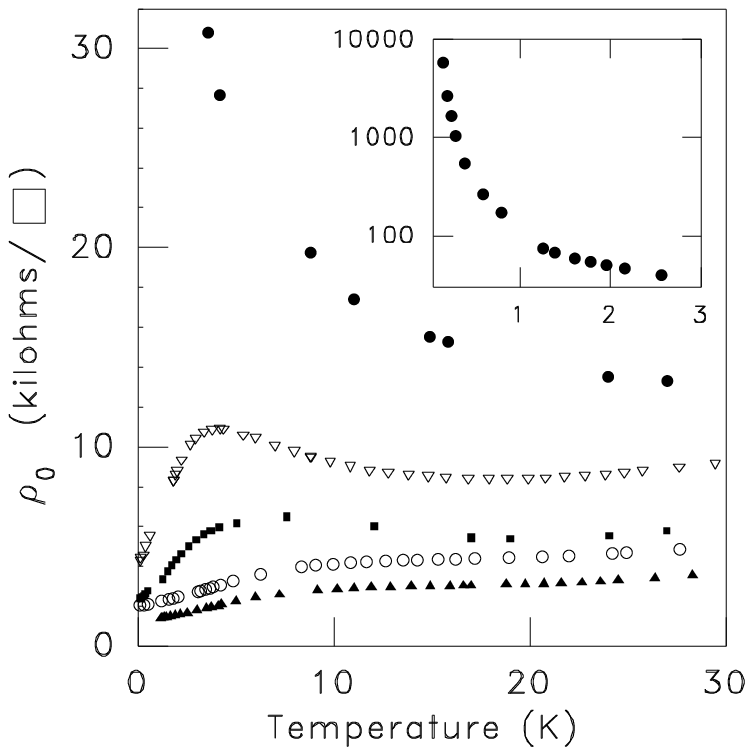}
         \large(f)\includegraphics[width=0.32\textwidth]{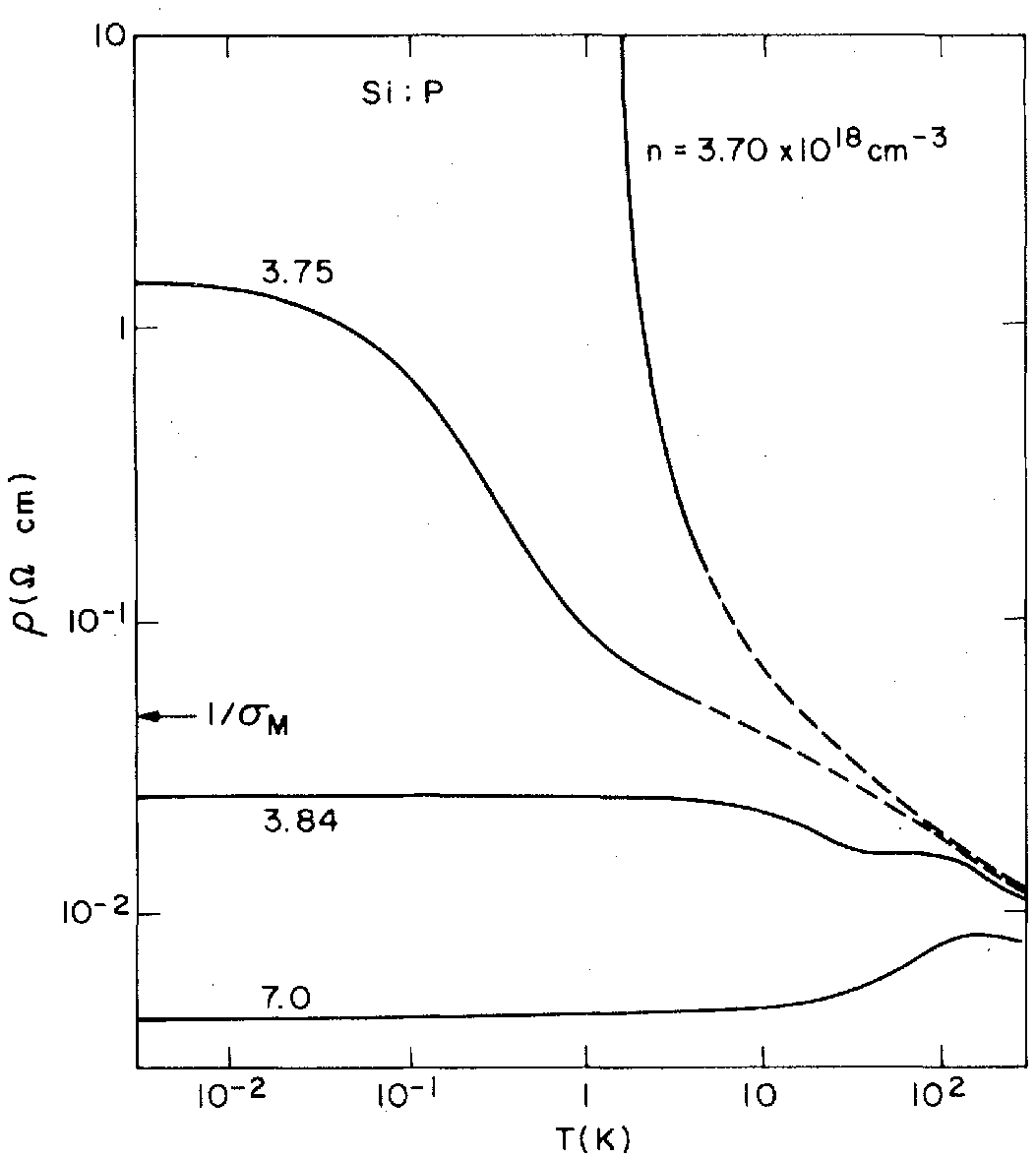}
    \caption{Critical behavior of the resistivity near the 2D metal-insulator transition in a Si MOSFET (a), p-GaAs/AlGaAs heterostructure (b), n-SiGe heterostructure (c), n-GaAs/AlGaAs heterostructure (d), and p-SiGe quantum well (e). The corresponding ranges of $r_s$ are $14-19$ (a), $12-32$ (b), $27-47$ (c), $5-14$ (d), and $7-19$ (e). For comparison, we also show the critical behavior of the resistivity in Si:P, a 3D system (f). Figures adapted from \textcite{krav}, \textcite{hanein98}, \textcite{lai05}, \textcite{eisenstein}, \textcite{coleridge97}, and \textcite{rosenbaum83}, respectively.}
    \label{fig:fig1}
\end{figure*}

\subsection{
Near  critical samples with $\rho \sim h/e^2$}
\label{metal-insulator tr}

Both the 2DEG and the 2DHG  exhibit a zero-temperature metal-insulator transition as a function of the electron density $n$.
This effect has been observed in all materials in which large $r_{\text s}$ samples can be made with sufficiently high mobilities: p- and n-Si MOSFETs, p- and n-GaAs quantum wells, p- and n- SiGe quantum wells, and n-AlAs quantum wells. Examples of the experimental data showing the resistivity $\rho(T)$ for different electron concentrations in various systems are presented in Fig.\ref{fig:fig1}.

  In all cases,  at low temperatures $T$, the resistivity  $\rho(T)$
exhibits a ``metallic'' temperature dependence ($d \rho(T)/d T>0$)
for electron concentrations $n$
in excess of a well defined critical value, $n_{\text c}$, and dielectric behavior ($d \rho(T)/d T<0$)
for $n < n_{\text c}$.  Moreover,  typically at the lowest temperatures $\rho \ll h/e^2$ for $n >n_{\text c}$,
while $\rho \gg h/e^2$ for $n < n_{\text c}$.
(Note that, were the Drude formula valid here, $\rho=h/e^2$ would correspond to a mean-free-path, $\ell$, equal to the Fermi wave-length: $k_{\text F}\ell=2\pi$.  The quantum of resistance, $h/e^2$, is thus the upper limit for the possible regime of applicability of Boltzmann transport theory.)

An issue that has been   debated is whether there is an actual transition or just a rapid crossover.  This issue is difficult to resolve unambiguously
 since a continuous metal-insulator transition is sharply defined only at zero temperature  as the point at which the resistance changes non-analytically from being finite (metallic) to infinite (insulating).  (Note, under the heading ``metallic'' we include the possibility of a ``perfectly metallic'' phase in which $\rho$ is non-zero at any non-zero $T$, but $\rho \to 0$ as $T\to 0$.)  However, at non-zero temperature, even where there is a zero temperature transition, at any 
 $T >0$, such a transition would manifest itself as a rapid, but analytic crossover. It is clear from Fig.\ref{fig:fig1} that the resistivity changes by many orders of magnitude as  $n$ varies over a modest range near $n_{\text c}$. The lower the temperature is, the more violent the resistivity changes.  For comparison,  in Fig.\ref{fig:fig1}~(f) we present data for the 3D metal-insulator transition in Si:P.  No one doubts that this data reflects a zero temperature metal insulator transition, although from a purely empirical viewpoint, the evidence is no better (and no worse) than in 2D.\\

 \begin{figure*}
         \large(a)\includegraphics[width=0.35\textwidth]{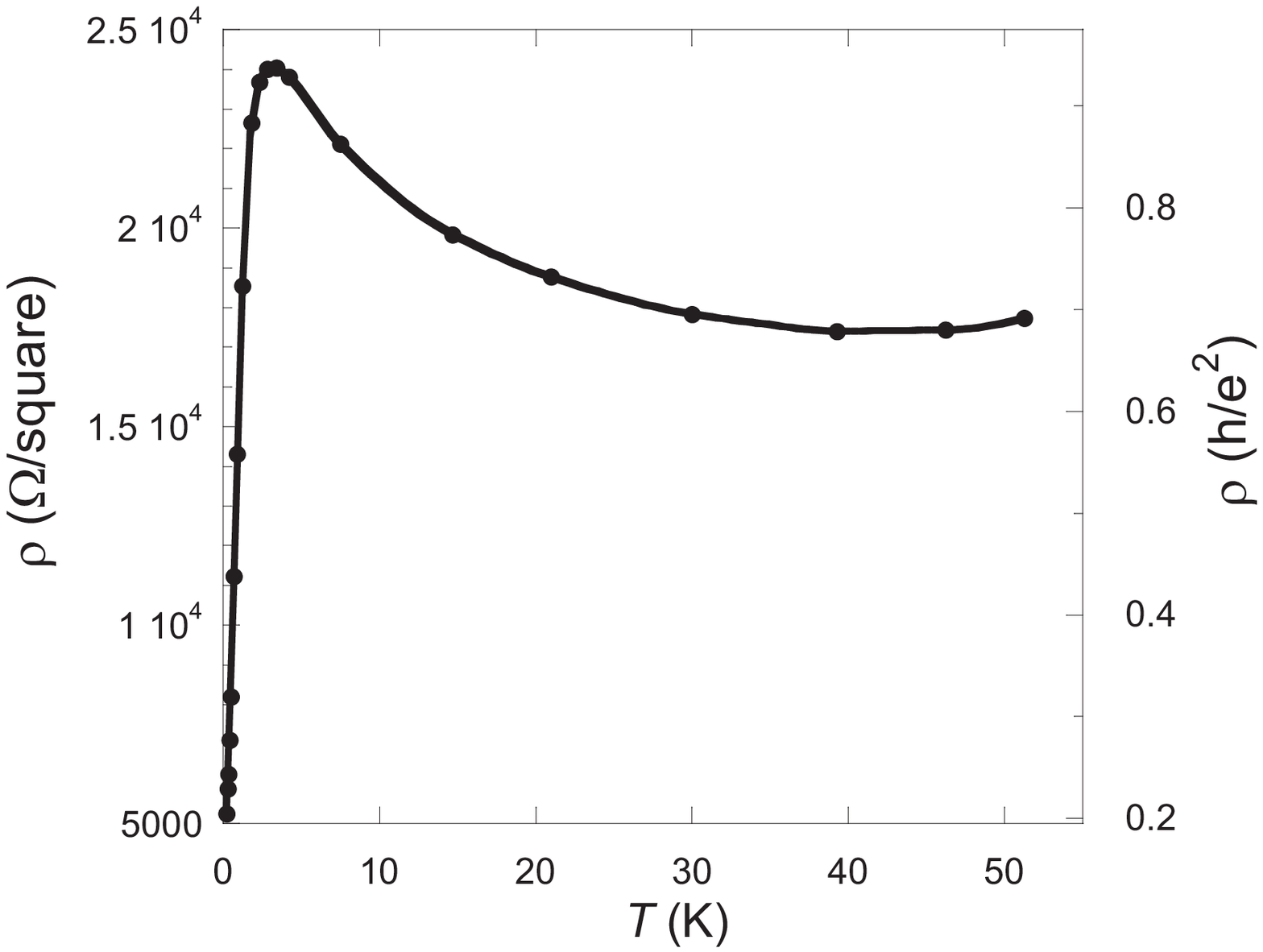}
        \large(b)\includegraphics[viewport=0 -15 600 650,width=0.438\textwidth,clip]{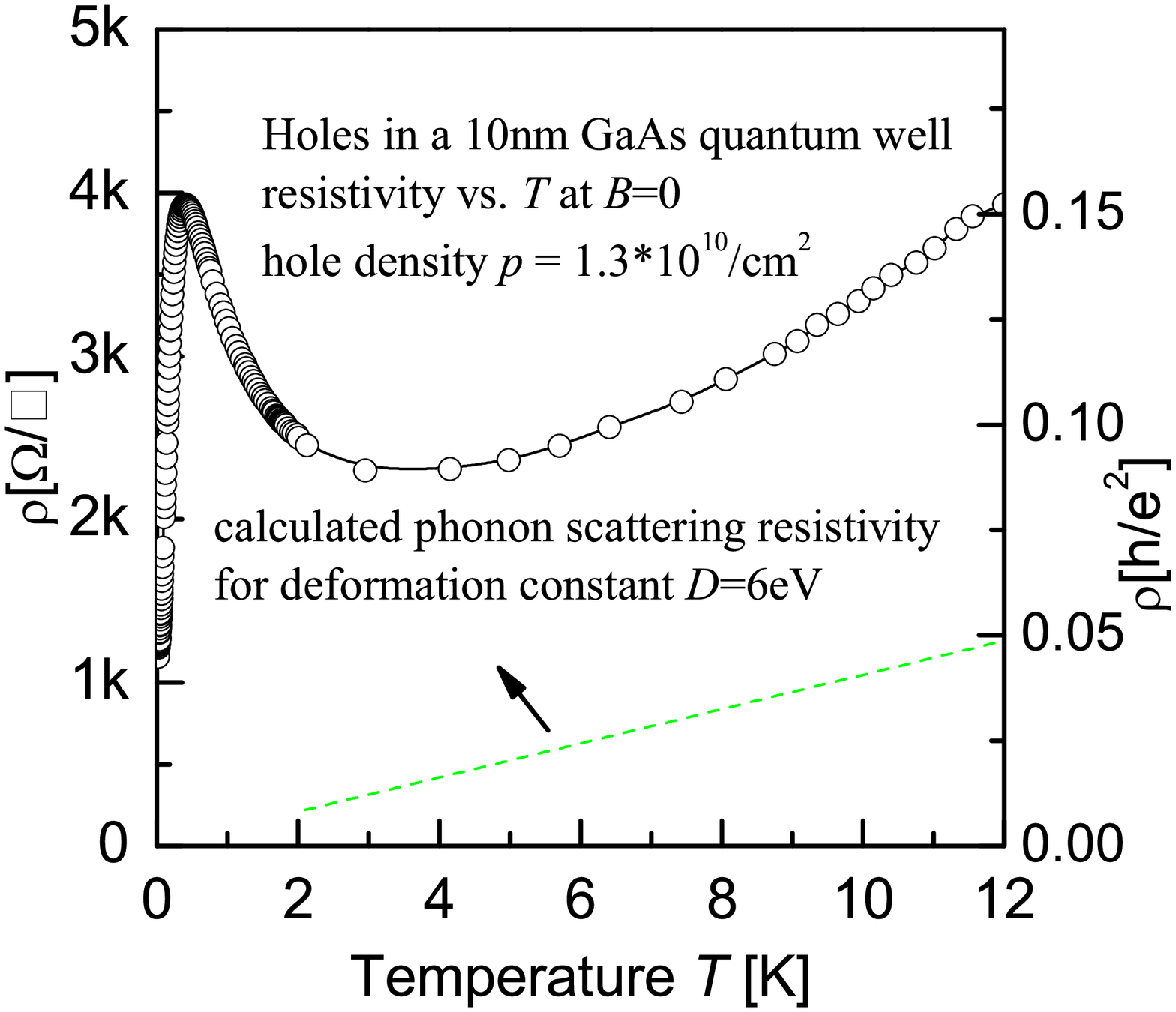}
         \large(c)\includegraphics[width=0.43\textwidth]{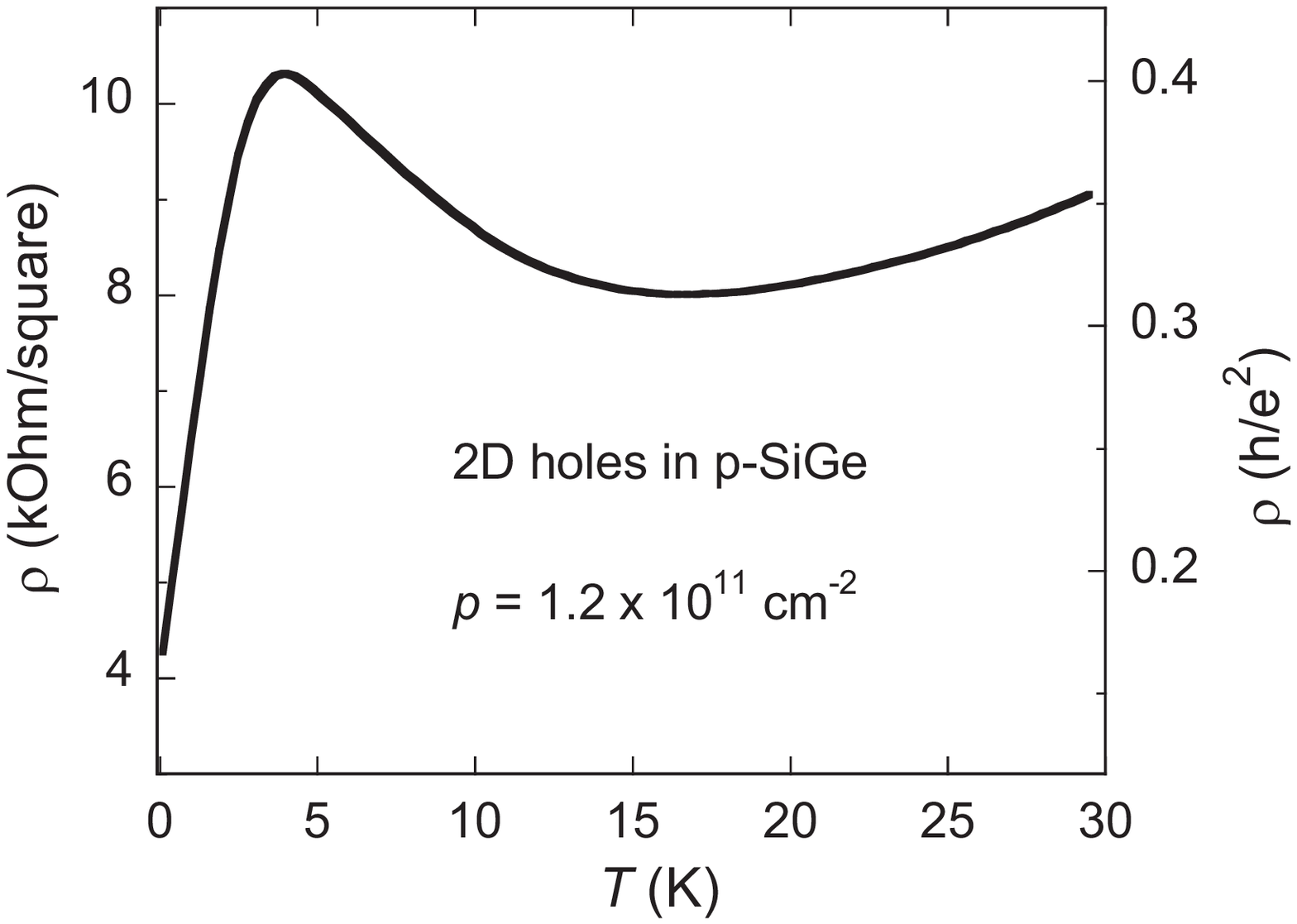}
         \large(d)\includegraphics[width=0.43\textwidth]{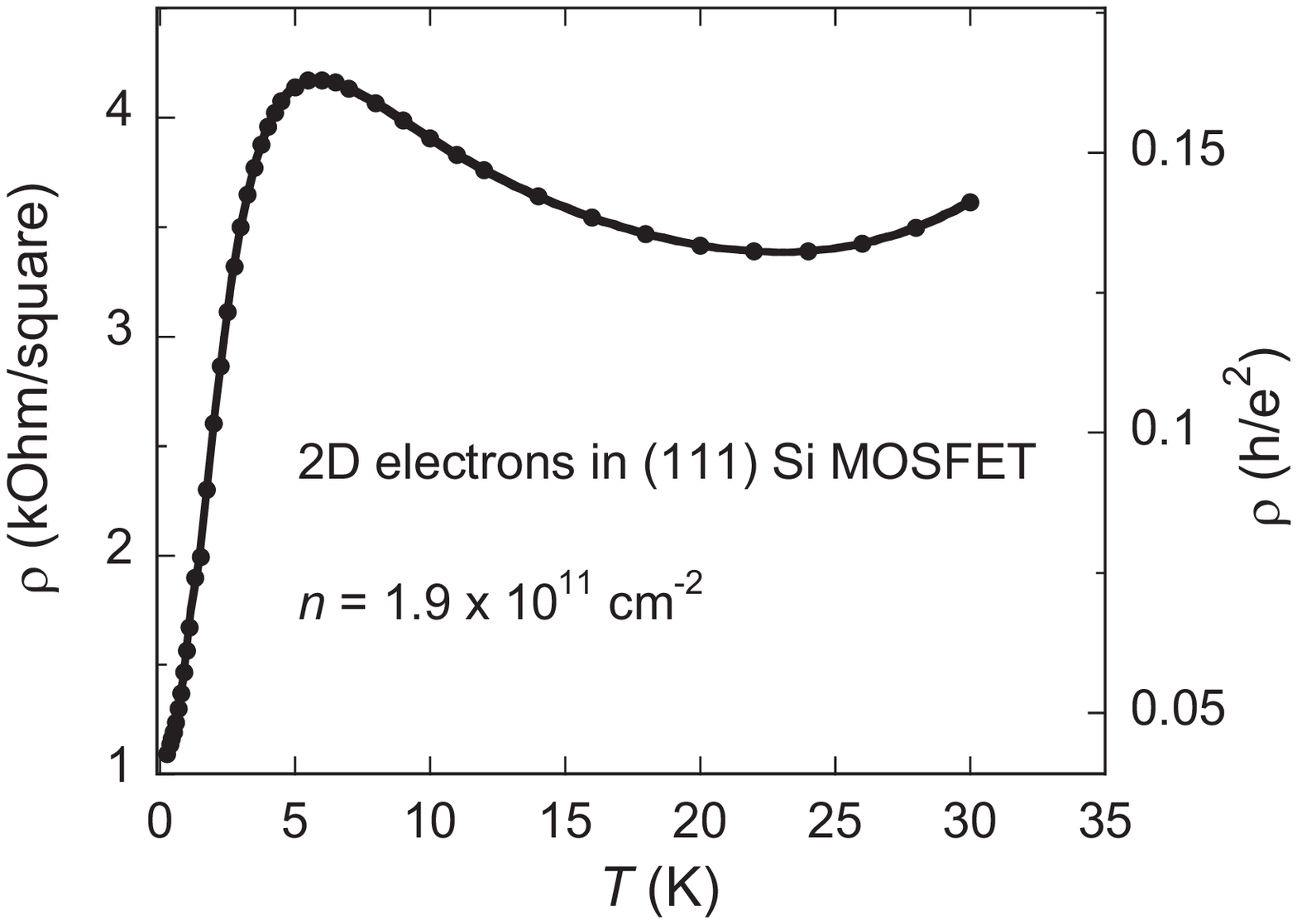}
    \caption{Non-monotonic temperature dependence of the resistivity in a (100) Si MOSFET (a), p-GaAs quantum well (b), p-SiGe quantum well (c), and (111) Si MOSFET (d) deep in the metallic regime over an extended temperature range. The bare (non-renormalized) Fermi temperatures are $7.5$~K (a), $0.75$~K (b), and $7$~K (c). Adapted from \textcite{krav_unp} (a), \textcite{Gao2005} (b), and
\textcite{coleridge_unp} (c). Panel (d) courtesy of
R.~N. McFarland and B.~E. Kane.}
    \label{fig:fig2}
\end{figure*}

We take the point of view that, at an empirical level, the experimental data presented in Fig.\ref{fig:fig1} 
 represent a metal-insulator transition in 2D.  It is, of course, possible that the non-analytic behavior
 would ultimately be rounded out if it were possible to follow the physics to much lower temperatures than
 have yet been attained, but this would necessarily involve a crossover to new physics.  However, the physics in the currently accessible range of temperatures is important to understand in its own right.

\subsection{
Strongly correlated highly metallic samples ($r_{\text s}\gg 1$ and $\rho\ll  h/e^{2}$):}
\label{StronlyCorrelatedMet}

In this section we focus on samples with $\rho \ll h/e^2$. In this limit, there is a small parameter, $1/k_{\text F}\ell \ll 1$, which permits a well controlled perturbative expansion of physical quantities.  Because the predictions of Fermi liquid theory are sharp,  discrepancies between experiment and theoretical expectations  can be readily documented.

\subsubsection{Temperature dependence of $\rho(T)$.}
\label{TempDep}

As shown in Figs.~\ref{fig:fig1}~(a-e) and \ref{fig:fig2}, in  samples  with large $r_{\text s}$ and $\rho\ll  h/e^{2}$, the resistance $\rho(T)$ is observed to increase with increasing temperature to a peak value $\rho(T_{\text {max}})$. Depending on which type of device and the value of $n$, the ratio $\rho(T_{\text {max}})/\rho(T_0)$ ranges from 2 to 10, where $T_0 \sim 25$~mK is the lowest temperature at which resistivity measurements are readily carried out.

This behavior has been observed in  many 2D electronic systems: Si
MOSFETs, p- and n-GaAs quantum wells, p- and n-SiGe quantum wells,
and in n-AlAs quantum wells. Even at the maximum, the resistance
$\rho_{max}=\rho(T_{\text {max}})$ is often smaller (and sometimes
much smaller) than the quantum of resistance. For instance, the
curves in Fig.\ref{fig:fig2}~(b), (c), and (d) and the corresponding
curves in Fig.\ref{fig:fig1}~(b), (c), (d), and (e) are deep in the
metallic regime. Moreover, generally it is found that $k_B T_{\text {max}} \sim E_{\text F}$, where $E_{\text F}$ is the bare Fermi energy.

As a function of increasing temperature when  $T > T_{\text {max}}$, the
resistance decreases, by as much as a factor of two (see Fig.\ref{fig:fig2}), before
ultimately starting to increase again at higher temperatures
where, presumably, scattering from thermally excited phonons
starts to be important. Still deeper in the metallic regime
({\it i.e.} at smaller $r_{\text s}$), the resistivity is a
monotonically increasing function of the temperature (see the lowest curves in Fig.\ref{fig:fig1}~(a-e)).

\subsubsection{Parallel field magneto-resistance.}
\label{ParallelFieldMR}

\begin{figure*}
    \centering
        \includegraphics[width=0.45\textwidth]{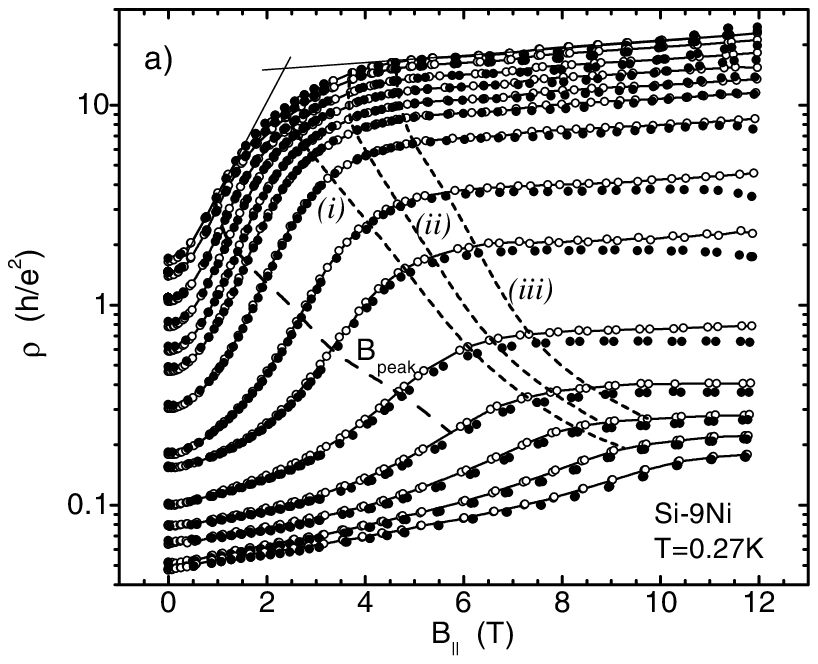}
        \includegraphics[width=0.43\textwidth]{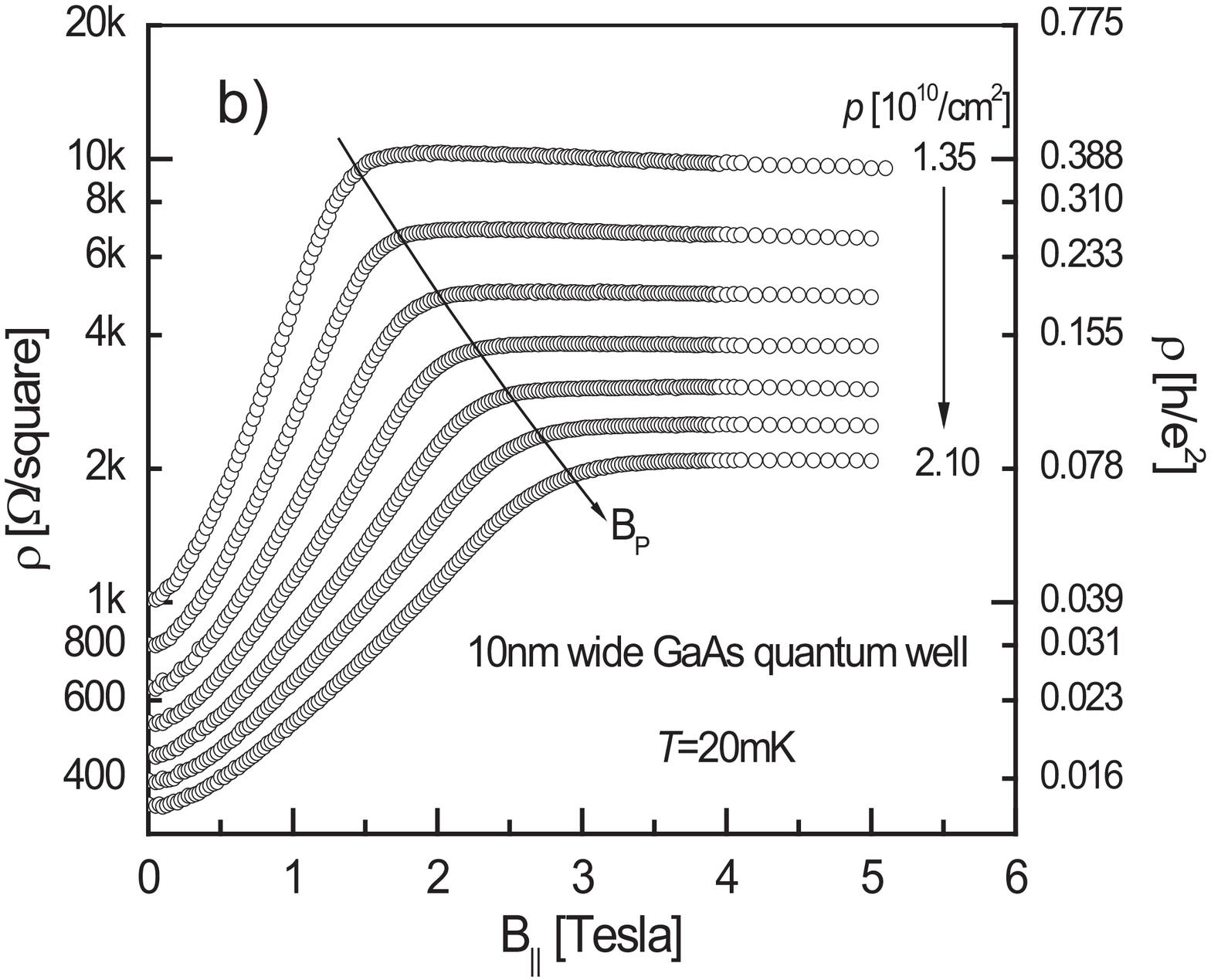}
        \includegraphics[viewport=0 0 532 523,width=0.4\textwidth,clip]{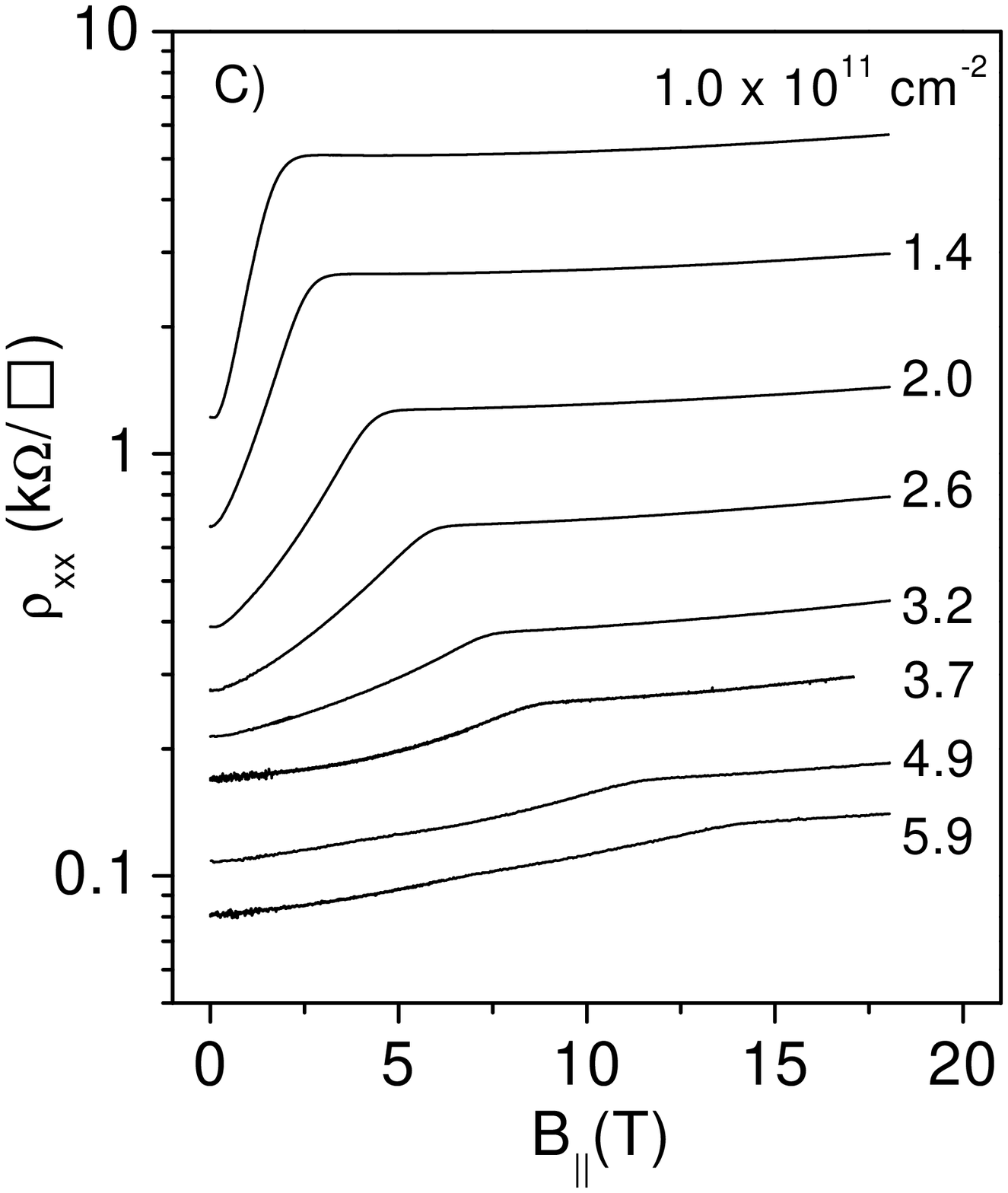}
        \includegraphics[viewport=0 0 291 617,width=0.25\textwidth,clip]{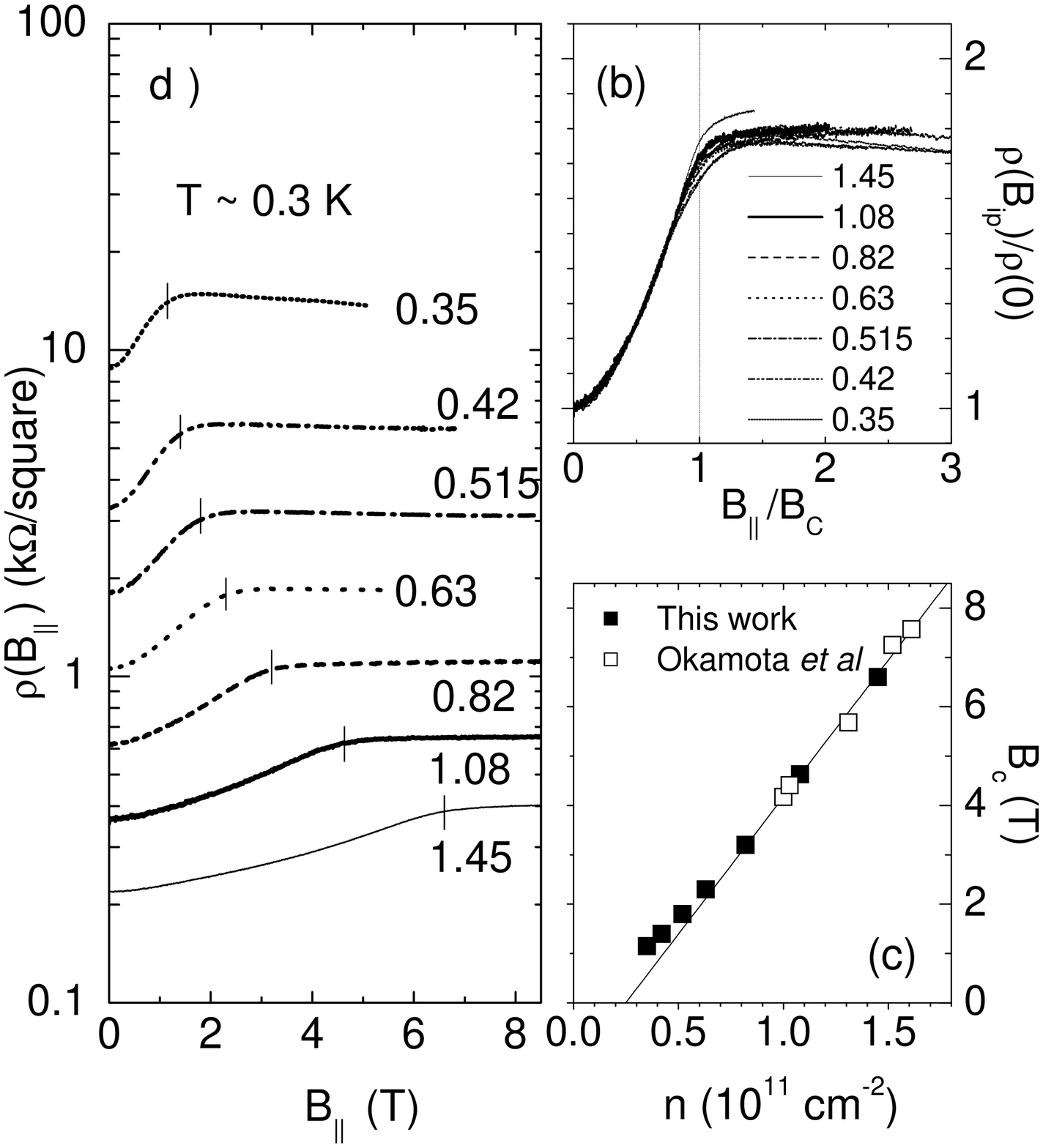}
    \caption{Giant magnetoresistance in a parallel magnetic field in a strongly metallic Si MOSFET (a), a 10 nm wide p-GaAs quantum well (b), a n-AlAs quantum well (c), and a n-SiGe quantum well (d). Adapted from \textcite{pudHpar}, \textcite{GaoHpar05}, \textcite{depoortere02} and \textcite{lai05}, respectively.}
    \label{fig:fig3}
\end{figure*}

As shown in Fig. \ref{fig:fig3},  in  Si MOSFETs,  p-GaAs heterojunctions, and SiGe quantum wells with large $r_{\text s}$ and $\rho\ll  h/e^{2}$, the resistance $\rho(T, B_{\|})$ for temperatures $T<E_{\text F}$ exhibits a strongly {\it positive} magneto-resistance, increasing by as much as an order  of magnitude as a function of $B_{\|}$, before saturating for $B_{\|}>B^{*}\sim E_{\text F}/g\mu_B$.  Here $\mu_B$ is the effective Bohr magneton,  $g$ is the gyromagnetic ratio, and $B_{\|}$ is the magnetic field parallel to the film. In sufficiently thin samples, $B_{\|}$ has little effect on the orbital motion of the electrons,  so  it can be viewed as coupling only to the electron spin,   and therefore the magneto-resistance is directly a function of the degree of spin polarization of the electron liquid.

\begin{figure}
    \centering
         \includegraphics[width=0.32\textwidth]{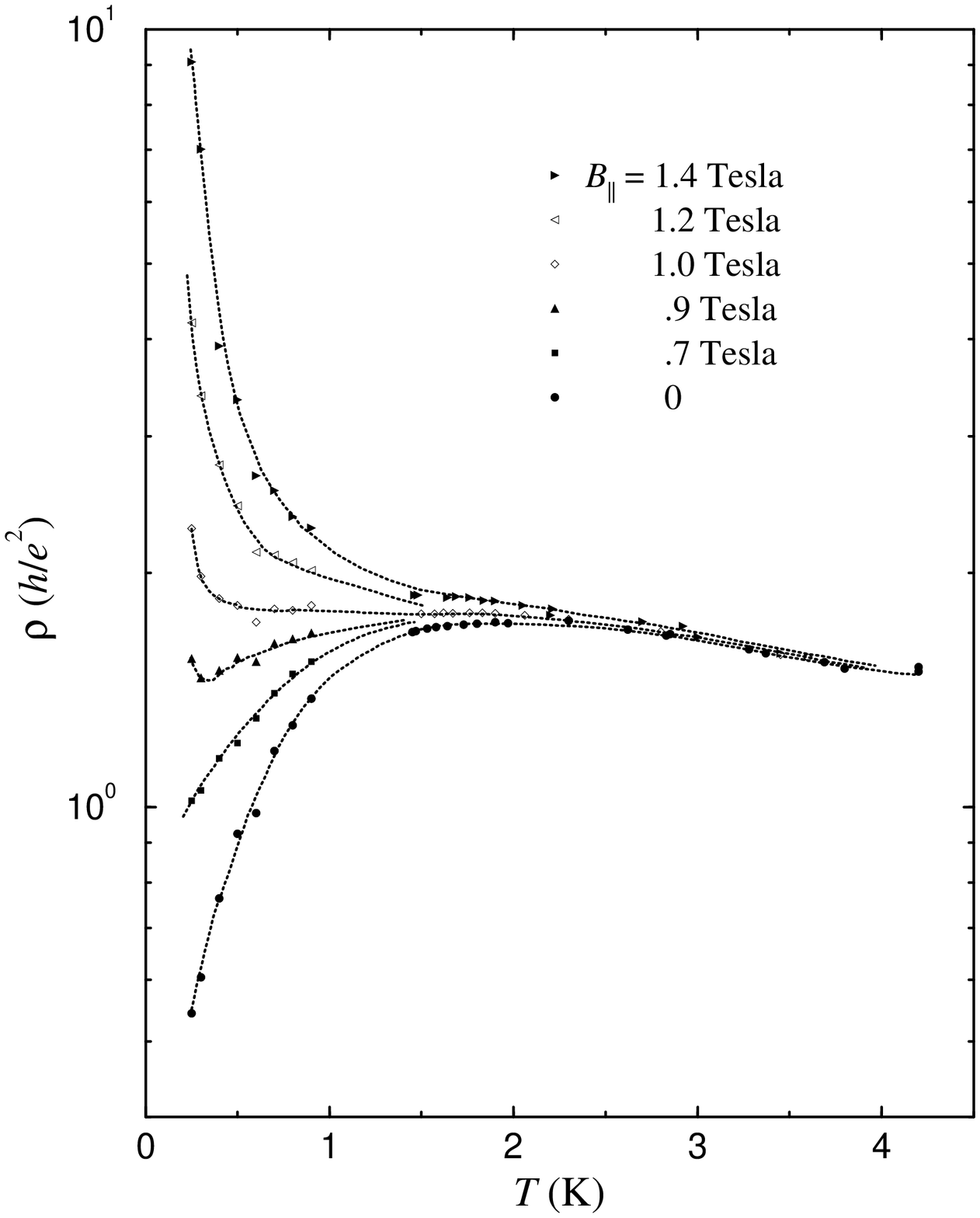}
    \caption{Parallel magnetic field-induced metal-insulator transition in a Si MOSFET. Adapted from \textcite{kravHpar}.}
    \label{fig:fig4}
\end{figure}

In samples that are sufficiently close to the point of the
metal-insulator transition, $B_{\|}$ can even induce a metal
insulator transition. This effect has been seen in  Si-MOSFETs
(see Fig.\ref{fig:fig4}).

\begin{figure}
    \centering
         \large(a)\includegraphics[width=0.40\textwidth]{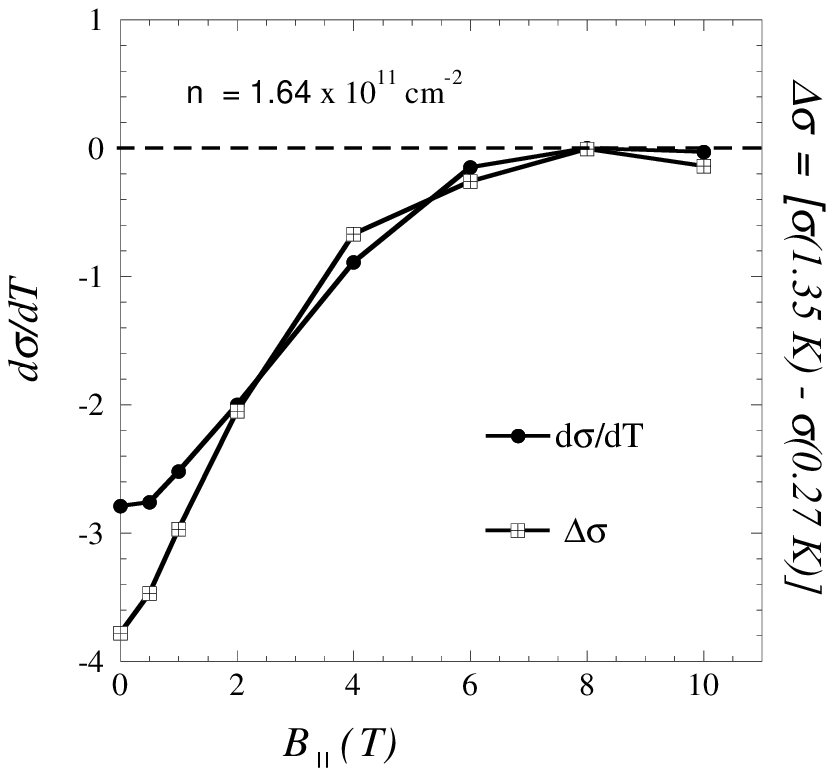}
        \large(b)\includegraphics[viewport=0 -6 490 400,width=0.40\textwidth,clip]{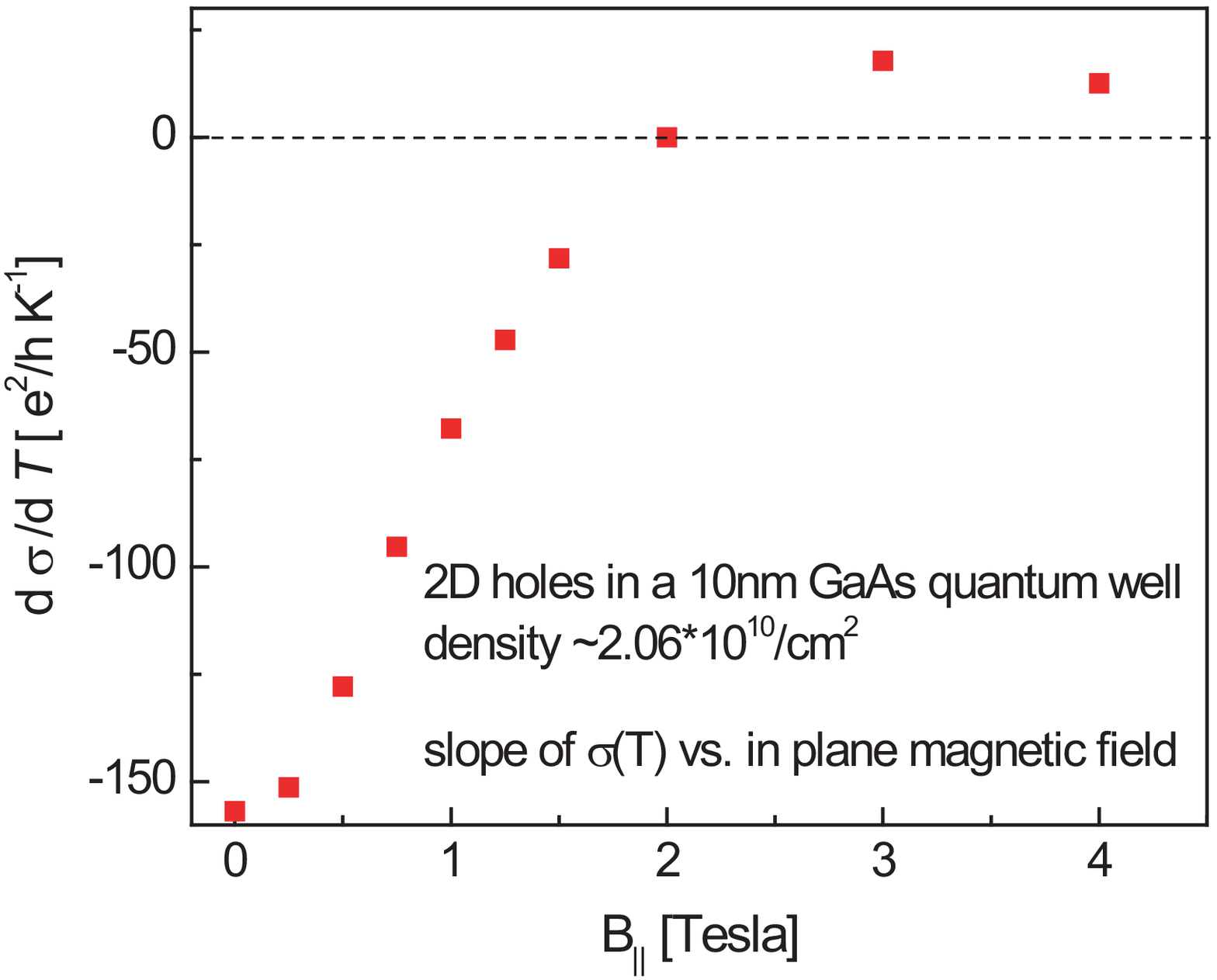}
    \caption{The slope $d\sigma/dT$ as a function of the parallel magnetic field in a Si MOSFET (a) and in a 10 nm wide p-GaAs quantum well (b). Adapted from \textcite{VitSat} and \textcite{GaoHpar05}, respectively.}
    \label{fig:fig5}
\end{figure}

As shown in Fig.\ref{fig:fig5} (a) and (b), the $T$ dependence of
$\rho$ at low temperatures ($T < T_{\text {max}}$) is largely eliminated
when the electron spins are polarized.  Specifically,
for magnetic fields $ B_{\|}> B^\star$, the slope
$d\rho(T,B_{\|})/dT$ is reduced from its $B_{\|}=0$ value  -- in
some cases by as much as two orders of magnitude! This effect
has been observed in Si MOSFET's and p-GaAs quantum wells.

\subsubsection{Magnetoresistance in a perpendicular magnetic field }
\label{MRperpMIT}

\begin{figure*}
    \centering
         \large(a)\includegraphics[width=0.40\textwidth]{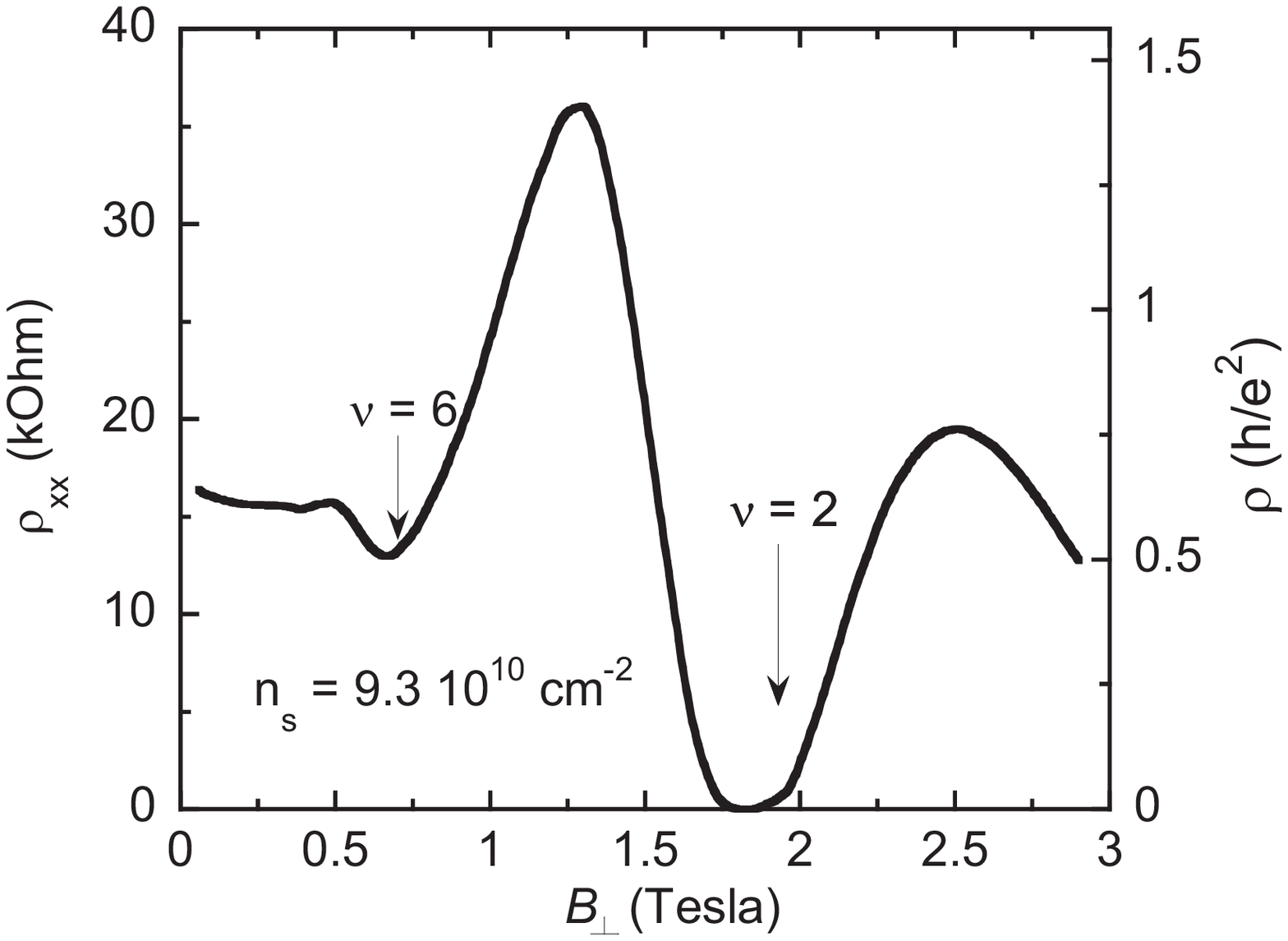}
         \large(b)\includegraphics[width=0.50\textwidth]{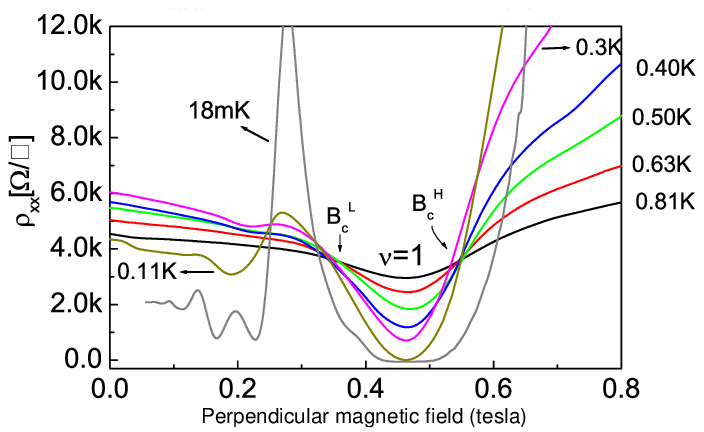}
    \caption{Magnetoresistance of the strongly correlated metallic 2D systems in a perpendicular magnetic field in a Si MOSFET (a) and a 10nm wide p-GaAs quantum well (b). Adapted from \textcite{kravchenko00a} and \textcite{GaoLanl}, respectively.}
    \label{fig:fig6}
\end{figure*}

The experimentally observed behavior of 2D strongly correlated
electron liquids in a perpendicular magnetic field $B_{\perp}$ can
be quite complex, as shown in Fig.\ref{fig:fig6}. In part, this
complexity reflects a combination of orbital and spin effects.

At small $B_{\perp}$ metallic samples often exhibit relatively small negative magneto-resistance. At larger $B_{\perp}$ the magneto-resistance is typically large and positive, while at still larger $B_{\perp}$ it becomes negative again as the system enters the quantum Hall regime.

The interplay between the quantum Hall states and the behavior of the system at large $r_{\text s}$ and $B_{\perp}\to 0$ has only been partially explored.
The phase boundary in the $n-B_{\perp}$ plane between the quantum Hall and insulating regimes
can be 
 identified in experiment 
 in at least two ways:  a)  Since the resistance is an increasing function of $T$ in the quantum Hall phase (vanishing as $T\to 0$) and a decreasing function in the insulating phase, the phase boundary can be approximately identified as the points at which $\rho_{xx}$ is $T$ independent.  b)  Since $\sigma_{xx}$ vanishes as $T\to 0$ in both the insulating and quantum Hall phases, but is non-zero in the critical regime,  the phase boundary can be approximately identified as the points at which $\sigma_{xx}$ has a  local maximum at the lowest accessible $T$.  At least when the field is not too small, these two methods produce essentially the same results, so the identification of the phase boundaries is unambiguous.

\begin{figure*}
    \centering
     \includegraphics[width=.7\textwidth]{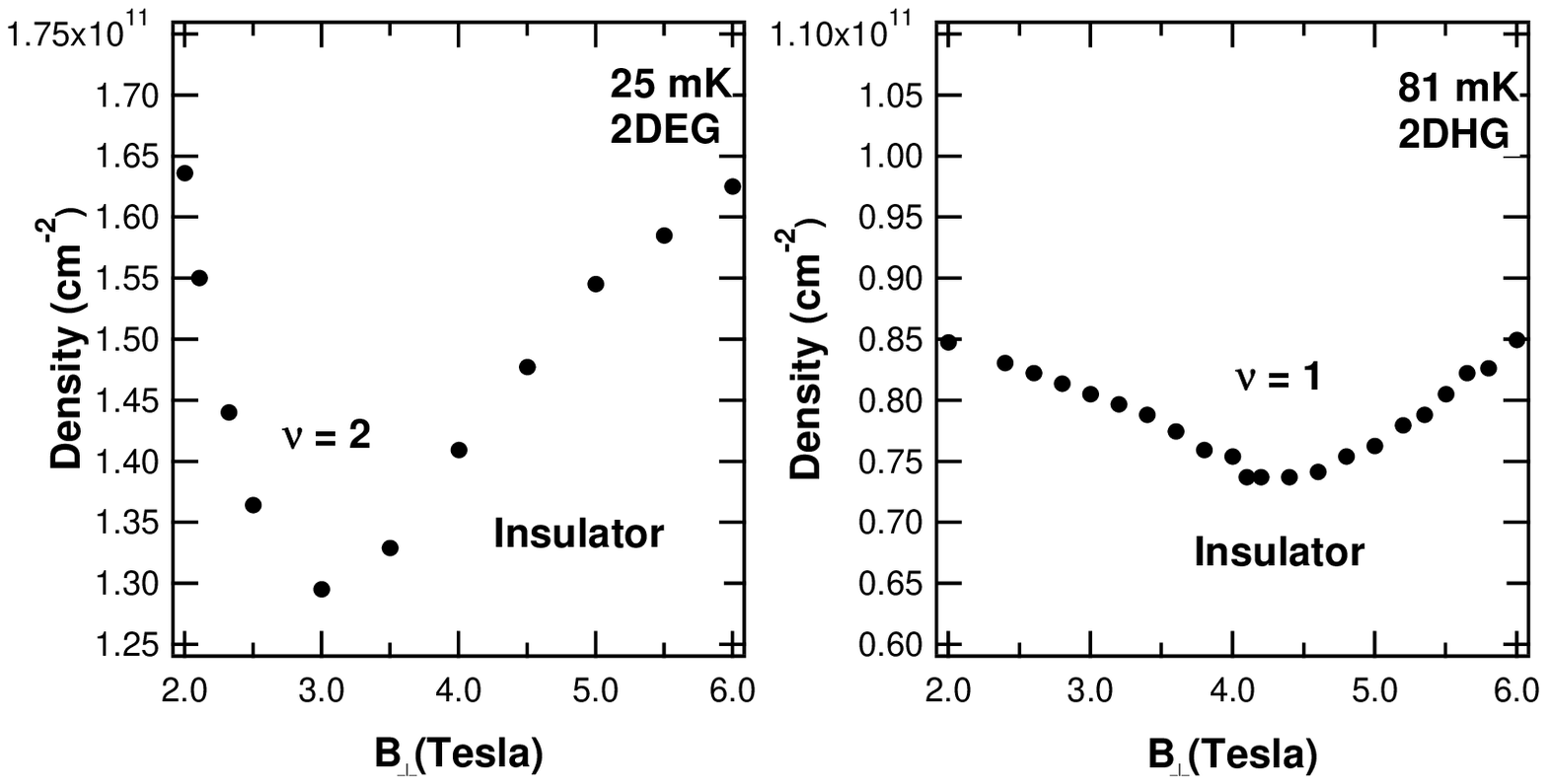}\vspace{5mm}\\
\caption{Position of the lowest extended state on a $n-B_{\perp}$ diagram in a ``weakly interacting'' ($r_s\sim1$) n-GaAs/AlGaAs heterostructure (left hand side panel) and in a ``strongly interacting'' ($r_s\sim10$) p-GaAs heterostructure (right-hand side panel). Adapted from \textcite{dultz98}.}
    \label{fig:fig7}
\end{figure*}

\begin{figure}
    \centering
         \includegraphics[viewport=0 0 580 407,width=0.45\textwidth,clip]{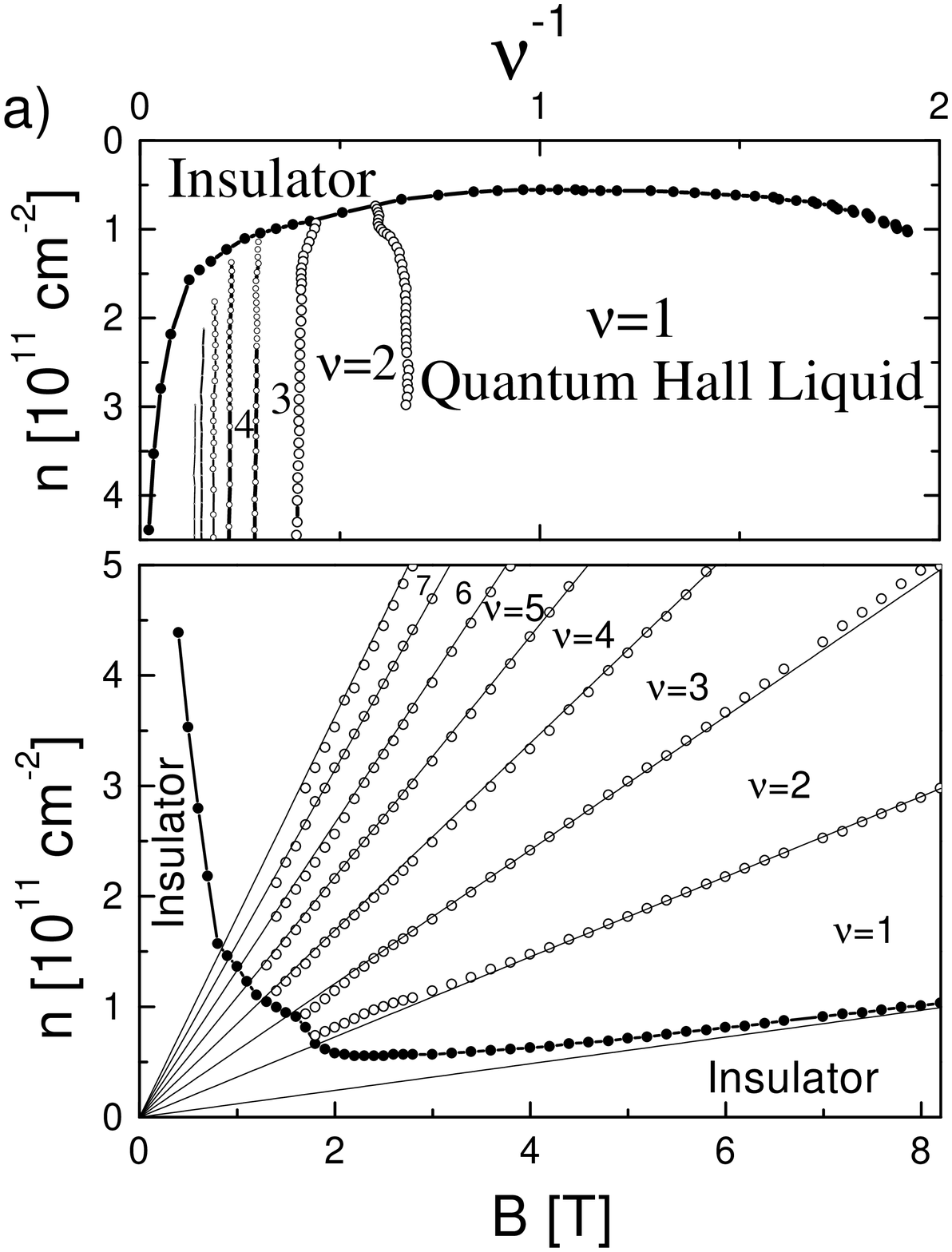}
    \caption{A map of the extended states for a highly-disordered 2D hole system in a Ge/SiGe quantum well. The open circles represent the positions of the extended state in the quantum Hall effect regime. The solid circles correspond to the position of the lowest extended state. Numbers show the value of $\rho_{xy}h/e^2$. Adapted from \textcite{hilke}.}
    \label{fig:fig8}
\end{figure}

 As shown in the left-hand side panel of Fig. \ref{fig:fig7}, in a n-GaAs sample with relatively small $r_{\text s}$, the phase boundary shifts up in energy as $B\rightarrow0$, presumably reflecting the expected ``floating'' of the extended states \cite{khmelnitskii84,laughlin84}
  ({\it i.e.} for non-interacting electrons, if all the states at $B_\perp=0$ are localized, then the lowest energy at which there is a localized state must diverge as $B_\perp \to 0$).  This ``floating'' can be seen even better in Fig. \ref{fig:fig8}, where the energy of the lowest extended state in a strongly-disordered 2D hole system in a Ge/SiGe quantum well is observed to increase by an order of magnitude as $B_\perp\rightarrow0$ \cite{hilke}.

\begin{figure}
    \centering
         \includegraphics[viewport=0 0 442 505,width=0.35\textwidth,clip]{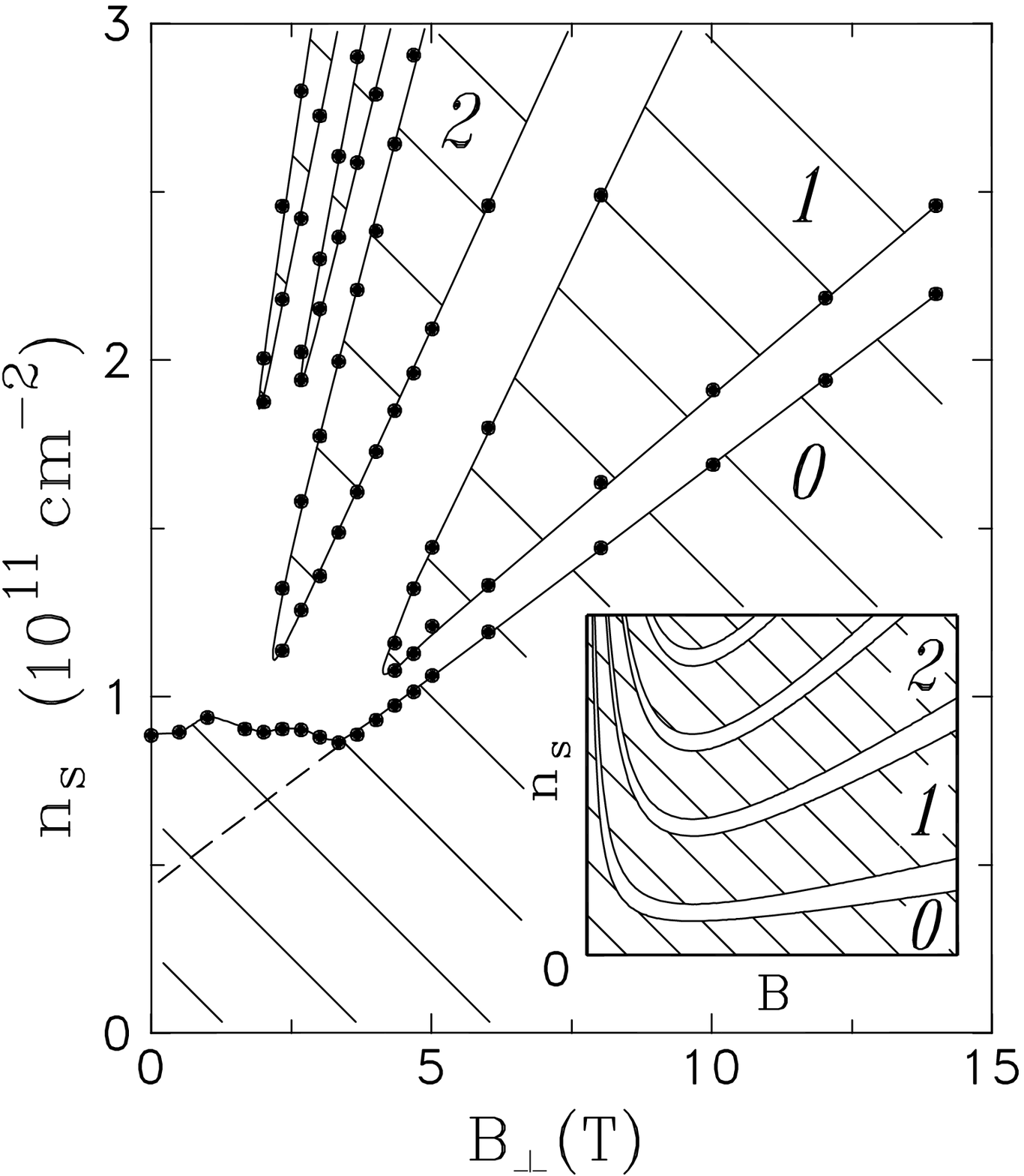}
\caption{Bands of the extended states (white areas) in a dilute Si MOSFET. The shaded area corresponds to localized states. The inset shows the expected ``floating'' of the extended states \cite{khmelnitskii84,laughlin84}. Adapted from \textcite{shashkin93}.}
    \label{fig:fig9}
\end{figure}

  In contrast, in high mobility p-GaAs samples with large $r_{\text s}$, as $B_{\perp} \to 0$ this phase boundary is observed to extrapolate to roughly the same value, $n_{\text c}$, which marks the zero field  metal-insulator transition in the same device \cite{dultz98}; see the right-hand side panel of Fig. \ref{fig:fig7}.  The analogous phase boundary, which also extrapolates to $n_{\text c}$, has been traced in Si MOSFETs \cite{shashkin93}, as shown in Fig. \ref{fig:fig9}. While there are some notable differences in the way, within the quantum Hall regime, the various different integer quantum Hall phases terminate at low fields, in both p-GaAs and Si MOSFETs, the phase boundary between the quantum Hall and insulating phases clearly extrapolates to a finite zero field critical value. This further corroborates the identification of $n_{\text c}$ as a critical point.

At low fields, or elevated temperatures, and small $r_{\text s}$, one
naturally observes Shubnikov-de Haas oscillations of
$\rho(B_{\bot})$ in sufficiently clean 2D systems.  Similar
oscillations   have been observed in both the Si MOSFETs and
p-GaAs quantum wells with large $r_{\text s}$ (see Fig. \ref{fig:fig6}).
However, for the large $r_{\text s}$ samples reported by
\textcite{GaoLanl}, these oscillations persist up to temperatures
which are comparable with the bare Fermi energy. It is important
to stress that FL theory not only predicts the existence of
magnetic oscillations with period inversely proportional to the
area enclosed by the Fermi surface, but it also predicts that the
amplitude of these oscillations should decrease in proportion to
$\exp[-2\pi k_BT/\hbar\omega_c]$ where the cyclotron energy,
$\omega_c=eB/mc$, in turn must be less than 
 $E_{\text F}$.

\subsubsection{ Spin magnetization of the electron gas at large $r_{\text s}$}
\label{SpinMag}

\begin{figure}
    \centering
         \includegraphics[width=0.40\textwidth]{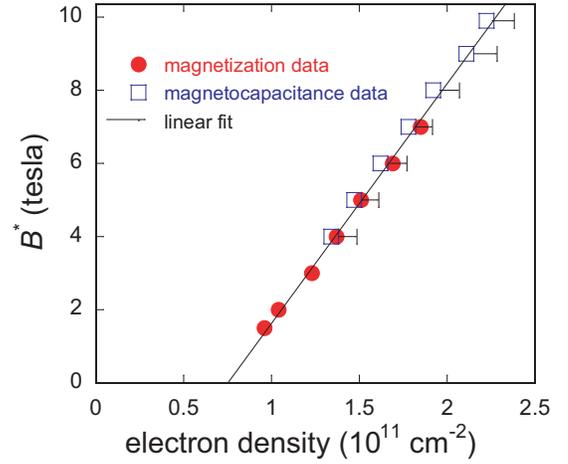}
    \caption{The critical magnetic field $B^*$ needed for complete spin polarization in a Si MOSFET as a function of the electron density. Adapted from \textcite{shashkin06}.  }
    \label{fig:fig10}
\end{figure}

The magnetization of the large $r_{\text s}$ metal is widely observed to exhibit a strong dependence on $n$, which reflects the increasing importance of
electron correlations. For instance, the in-plane magnetic field of complete spin polarization, taken from magnetization and magnetocapacitance measurements in clean Si-MOSFETs \cite{shashkin06},  decreases significantly more strongly with decreasing $n$ than does $E_{\text F}\propto n$, as shown in Fig.~\ref{fig:fig10}.  Strong dependences of the magnetization on $n$ have also been seen (\textcite{vitkalov01a}; \textcite{PudalovSpin}; \textcite{StormerZhuSpin}; \textcite{ShayeganSpin}; \textcite{TsuiSpin}) in other types of devices with $n$ near the critical density for the metal-insulator transition.  However, there are subtle, but important differences in the $n$ dependences.  For instance, in Si-MOSFETs, $B^*$ appears to extrapolate to 0 at a positive value of $n=n^*$ (\textcite{shashkin01a}; \textcite{vitkalov01a}; \textcite{shashkin06}), while in the Si/SiGe quantum well devices studied by \textcite{TsuiSpin}, the $n$ dependence of $B^*$ over the accessible range of $n$ was fitted by $B^* \sim n^{1.3}$, which vanishes only as $n\to 0$.

Complex evolution of the linear susceptibility, $\chi$, with $n$ has also been observed, but in contrast with the non-linear response (as parameterized by $B^*$), the linear response can be strongly affected by rare regions of localized spins which make a large contribution to the low $T$ susceptibility \cite{prus}. Nevertheless, experiments \cite{shashkin06} show that the $n$ dependent trends of $\chi$ are roughly in agreement with the results extracted from measurements of $B^{*}(n)$.

\subsection{Strongly correlated highly resistive  samples, $r_{\text s} \gg 1$ and $\rho > h/e^2$}
\label{HighRes}

The properties of the 2DEG at  large $r_{\text s}$ well on the insulating side of the metal insulator transition have been less completely explored, experimentally.  In comparing results from different systems, one vexing issue is to what extent the large values of $\rho$ reflect 
strong effects of disorder, as opposed to the intrinsic effects of correlations.  We are primarily interested in data on the cleanest possible systems, where at the very least the correlation effects must have strongly enhanced the effects of the weak disorder.

Here, we focus on some astonishing experimental observations in strongly correlated, highly resistive devices.

\subsubsection{
``Metallic'' $T$ dependence in samples with resistivity $\rho > h/e^2$}
\label{MetTdepInsValue}

\begin{figure}
    \centering
         \includegraphics[width=0.40\textwidth]{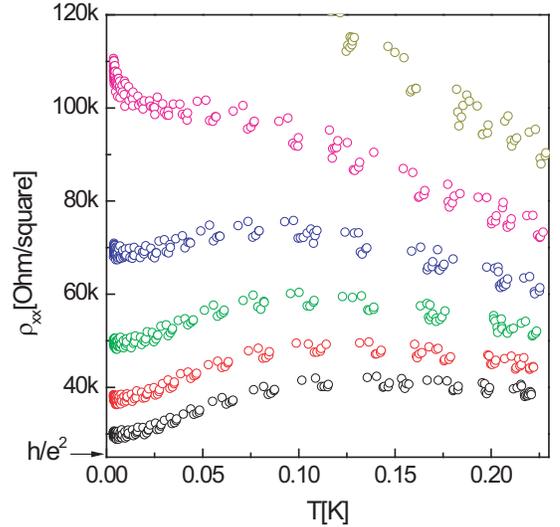}
    \caption{Metallic ($d\rho/dT>0$) temperature dependence of the resistivity in a 30 nm wide p-GaAs quantum well with $\rho>h/e^2$ (see four lower curves).  2D hole density is 6.1, 5.9, 5.7, 5.5, 5.3 and 5.1 $\times$10$^{9}$cm$^{-2}$ from bottom to top curve. Adapted from \textcite{gao_thesis}.}
    \label{fig:fig11}
\end{figure}

 The conventional theory  of localization in the strong disorder limit $\rho \gg h/e^2$ predicts that the
  electronic transport should be due to hopping conductivity, and hence $\rho$ should diverge strongly (exponentially) as $T\to 0$.
On the other hand, in  p-GaAs samples  with large $r_{\text s}$, as shown in Fig.\ref{fig:fig11}, the low temperature resistivity for a range of $n$ in which $\rho$ is up to 3 times larger than $h/e^2$ sometimes exhibits a ``metallic'' temperature  dependence, {\it i.e.} $\rho$ {\it increases} with
increasing temperature. Similar behavior has also been observed in Si MOSFETs (see the two middle curves in Fig.\ref{fig:fig1}~(a)).

 \subsubsection{ Magneto-resistance  for $r_{\text s}\gg 1$ and $\rho > h/e^{2}$.}
 \label{Parallel field MR}

\begin{figure}
    \centering
         \includegraphics[width=0.35\textwidth]{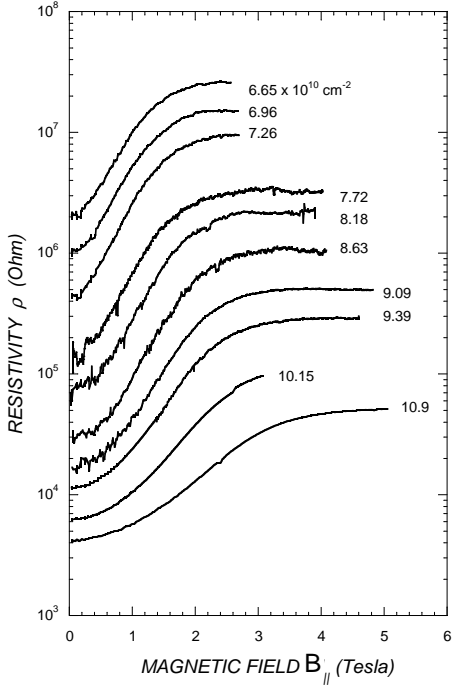}
    \caption{Parallel field magnetoresistance in a Si MOSFET at different electron densities across the metal-insulator transition. Adapted from \textcite{mertes99}.}
    \label{fig:fig12}
\end{figure}

The magneto-resistance   in a parallel magnetic field $B_{\|}$  on the insulating side of the metal-insulator transition is still large and positive, even in samples with $\rho$ 
 as high as 2 M$\Omega$/square.
 This can be seen in the data from Si MOSFET devices  shown in Fig. \ref{fig:fig12}.  At large enough $B_{\|}$ the magnetoresistance saturates in a way similar to that in metallic samples.

\begin{figure}
    \centering
     \includegraphics[width=0.40\textwidth]{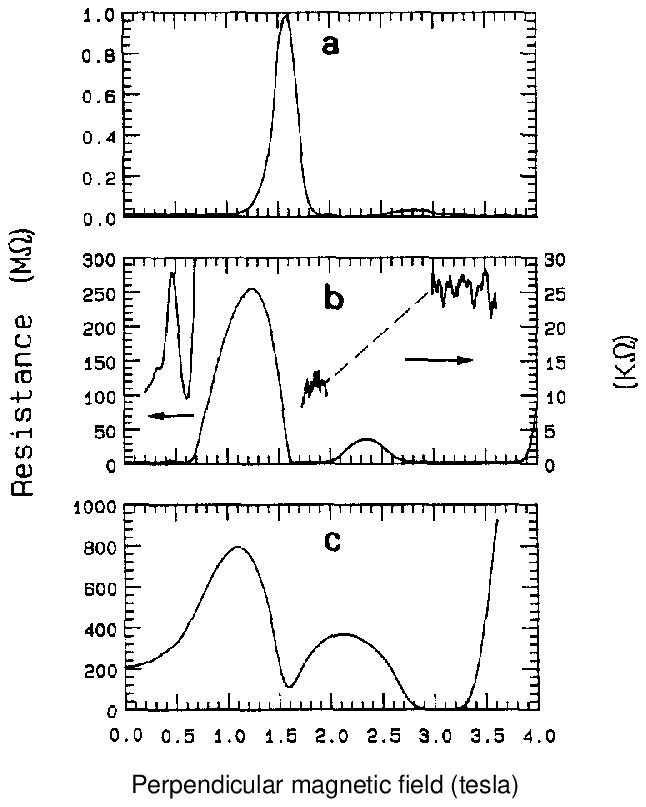}
        \includegraphics[viewport=0 -10 265 209,width=0.40\textwidth,clip]{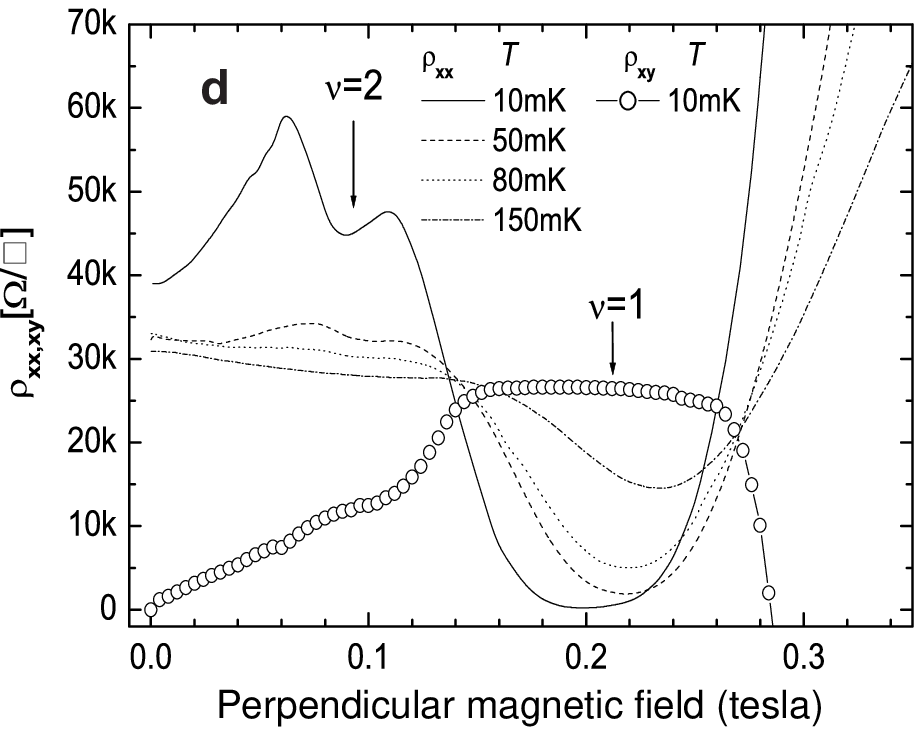}
    \caption{Magnetoresistance of a highly resistive dilute 2D gas in a perpendicular magnetic field in a Si MOSFET (a-c) and in a 30nm wide p-GaAs quantum well (d). Adapted from \textcite{diorio90} and \textcite{gao_thesis}, respectively.}
    \label{fig:fig13}
\end{figure}

 In a perpendicular magnetic field, $B_\bot$, samples with $\rho(B=0)\gg h/e^2$  exhibit rather complex magnetoresistance. As can be seen in Figs. \ref{fig:fig13}~(a-c), the resistance first increases strongly for small $B_\bot$, sometimes by orders of magnitude. At higher $B_\bot$, the magnetoresistance turns negative and eventually the system exhibits giant resistance oscillations with deep minima at magnetic fields corresponding to Landau level filling factors $\nu=1$ and $2$. The same general features have been observed in p-GaAs samples \cite{gao_thesis}, as shown in Fig.\ref{fig:fig13}~(d)).

\begin{figure}
    \centering
        \includegraphics[viewport=123 0 788 567,width=0.4\textwidth,clip]{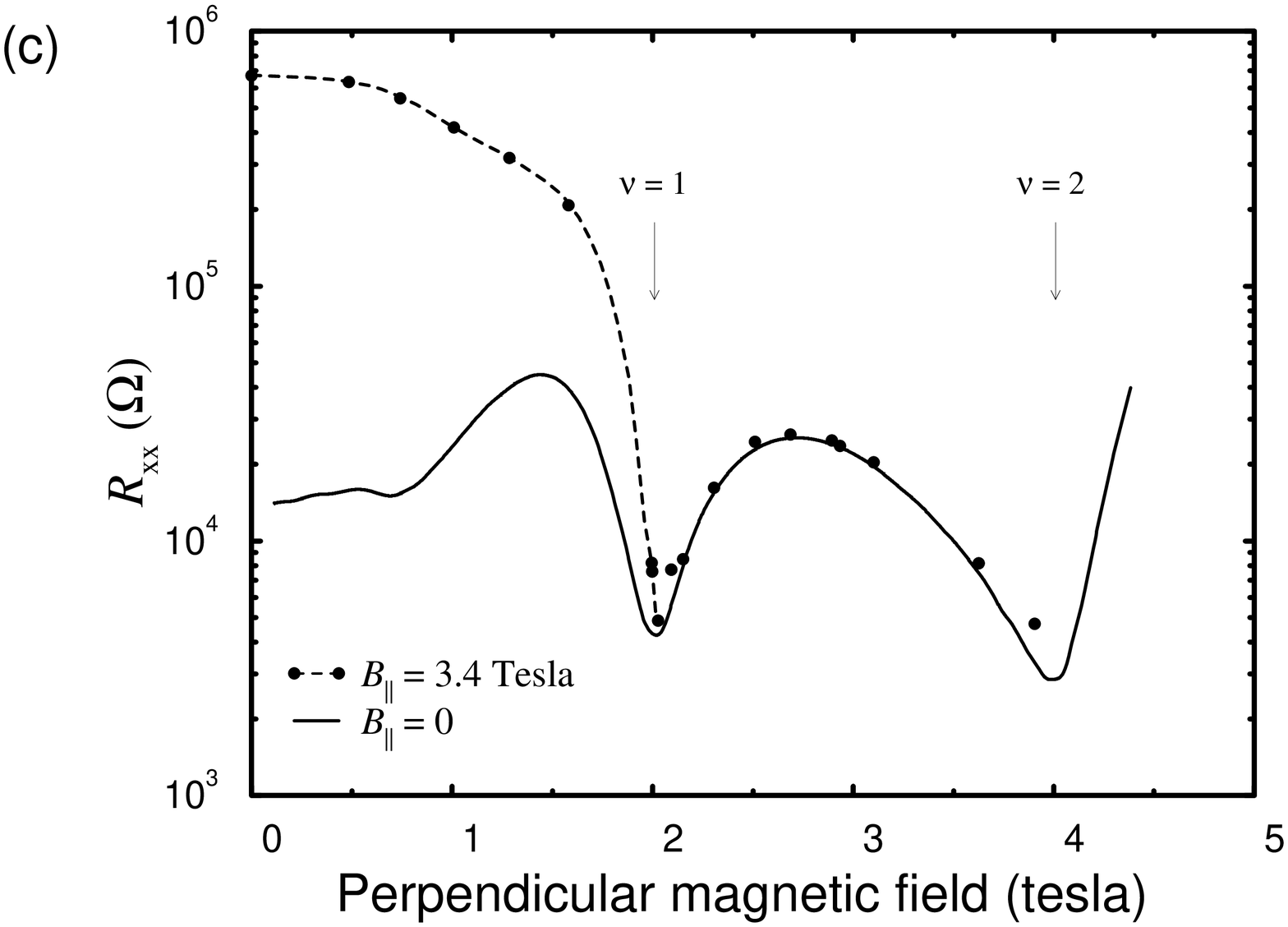}
    \caption{$R_{xx}(B_\perp)$ in a silicon MOSFET in the presence of a parallel field $B_{\|}=3.4$ tesla used to suppress the metallic behavior (solid symbols). For comparison, the magnetoresistance in the absence of a parallel magnetic field is shown by the solid line. Adapted from \textcite{kravchenko98}.}
    \label{fig:fig14}
\end{figure}

That the positive contribution to the magneto-resistance is largely a spin effect can be seen in Fig.  \ref{fig:fig14}, where, in the presence of a strong in-plane field, $B_{\|} \sim B^\star$, the resistance at $B_\perp=0$ is much larger than for $B_{\|}=0$, but then, as a function of increasing $B_\perp$, the magneto-resistance is everywhere strongly negative all the way to the quantum Hall regime.

\subsection{Drag experiments on double-layers with $r_{\text s} \gg 1$ and $\rho \ll h/e^2$}
\label{DragExp}

Additional information concerning correlation effects can be obtained from measurements of the ``drag''
 resistance in  a system of two 2DEG layers which are electrically unconnected.
Current
$I$ is passed through the lower (active) layer and  the voltage $V_{D}$ is measured on
the upper (passive) layer.
The drag resistance is defined to be the ratio $\rho_{D}=V_{D}/I$.

For relatively small $r_{\text s}$, measurements \cite{Eisenstein93,Eisenstein94} of $\rho_{D}(T)$ in
double layer 2DEGs are in qualitative agreement with Fermi liquid
theory \cite{Price,MacDonnald}. Specifically, the drag resistance is small,
 proportional  to $(T/E_{\text F})^2$ and 
  to $(k_Fd)^{-\alpha_d}$, where $k_{\text F}$ is
  the Fermi momentum, $d$ is the spacing between the two layers, and typically
   $\alpha_d = 2$ or 4, depending on the ratio of $d/\ell$.

\begin{figure}
    \centering
        \includegraphics{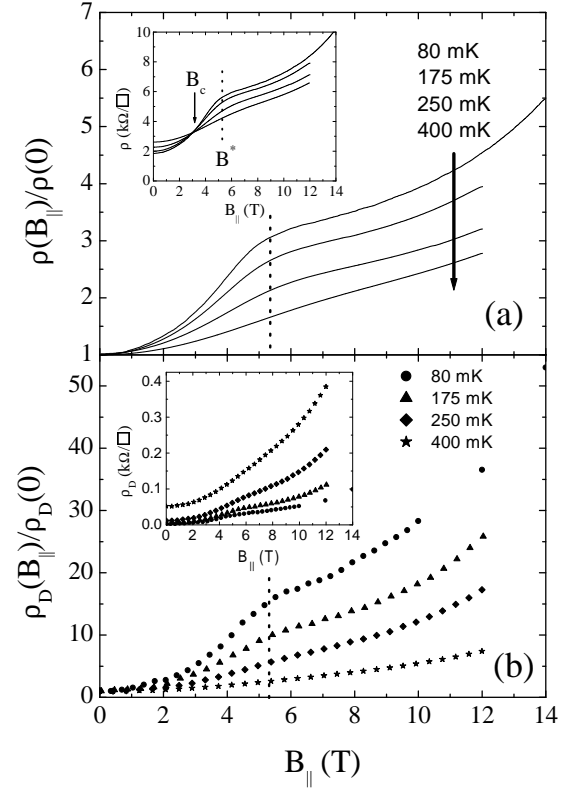}
    \caption{In-plane magnetotransport data for $p_m = 2.15\times10^{10}$
cm$^{-2}$ at $T$ = 80, 175, 250, and 400 mK. (a) Inset: $\rho$ vs $B_{\|}$.
Main Plot: Data from inset normalized by its $B_{\|}= 0$ value. (b) Inset: Corresponding data for $\rho_D$ vs $B_{||}$.
Main Plot: Data from inset normalized by its $B_{\|}= 0$ value. Adapted from \textcite{Pillarisetty}.}
    \label{fig:fig15}
\end{figure}

   However,   experiments
    \cite{Pillarisetty}  on p-GaAs double layers with $r_{\text s} \sim 20-30$
 yield results which differ significantly from 
 what would be expected on the basis of a simple extrapolation of the small $r_{\text s}$ results.
  These experiments were performed on samples with small resistances $\rho\sim 0.05 - 0.1\  h/e^2$,
in which quantum interference corrections to the Drude conductivity are presumably insignificant.
The experimental data shown in Fig. \ref{fig:fig15}
reveal the following features:
\begin{itemize}
\item  The drag resistance in these samples
is 1-2 orders of
magnitude larger than expected on the basis of 
a simple extrapolation of the small $r_{\text s}$ results.
\item    Whereas in a Fermi liquid, $\rho_{D}(T)\sim T^2$,
 in large $r_{\text s}$ devices $\rho_{D} \sim (T)^{\alpha_T}$ where
the temperature exponent  exhibits non-Fermi liquid values
$2 < \alpha_{T} < 3$.
\item At low temperature, $\rho_{D}(T,B_{\|})$ {\it increases} as a function of $B_{\|}$ by a
factor of 10-20  and saturates when $B_{\|}>B^{*}$.
A parallel magnetic field also appears to strongly suppress the temperature dependence of
$\rho_{D}(T)$.
In the presence of a non-zero $B_{\|}$, the value of
$\alpha_T(B_{\|})$ decreases
 with $B_{\|}$
 and saturates for $B_{\|}>B^{\star}$ at a value which is significantly
  smaller
than the FL value $\alpha_T=2$.
\item  The $T$ and especially the $B_{\|}$ dependences
  of $\rho_{D}(T, B_{\|})$ and the resistivities of the individual
   layers $\rho(B_{\|}, T)$ look
qualitatively similar to one another, which suggests that both have a common
    origin (see Fig. \ref{fig:fig15}). (In a Fermi liquid, $\rho(T, B_{\|})$ is primarily determined by the
 electron-impurity scattering, and $\rho_{D}(T, B_{\|})$ by the inter-layer
 electron-electron scattering, so there is no \emph{a priori} reason for their $T$ and
  $B_{\|}$ dependences to be similar.)
\end{itemize}

\subsection{Comparison with small $r_{\text s}$ devices}

 The anomalies discussed above have been observed  in samples with large $r_{\text s}$. In relatively smaller $r_{\text s}$ high mobility devices ( {\it i.e.}, $r_{\text s} \sim 1$), the observed behavior is much more in line with the expectations of FL theory, modified by weak interference effects.  For example, there is  a smooth finite temperature crossover as a function of disorder from a ``weak localization'' regime for $\rho < h/e^2$ to variable range hopping behavior for $\rho > h/e^2$; $\rho$ is weakly (logarithmically) dependent on $T$  for $\rho < h/e^2$ and $T\ll T_F$; the parallel field magneto-resistance is weak (logarithmic) and positive for $\rho < h/e^2$; the resistance diverges strongly (exponentially) with decreasing $T$ whenever $\rho > h/e^2$; the drag resistance is small compared to $\rho$ (see, for example, \textcite{stern} for a review).  Moreover, in the presence of a perpendicular magnetic field, small $r_{\text s}$ devices exhibit ``levitation of the delocalized states'', {\it i.e.}, the phase boundary between the quantum Hall and the insulating phases tends toward ever higher values of the density as $B_{\perp} \to 0$, as shown on the left hand side panel of Fig.~\ref{fig:fig7} and in Fig.~\ref{fig:fig8}.

\section{Theoretical Considerations}

\subsection{Good 2D ``metals'' with $r_{\text s} \ll 1$ and $k_{\text F}\ell \gg 1$}
\label{classicaltheory}
The properties of interacting electronic systems  crucially depend  on the dimensionless strength of the interactions, $r_{\text s}$, and the strength of the quenched disorder, which is parameterized by the dimensionless quantity $k_{\text F}\ell$, where $k_{\text F}$ is the electron Fermi momentum and $\ell$ is the electron elastic mean free path.
The theory of pure ($k_{\text F}\ell\gg 1$) and weakly interacting ($r_{\text s}\ll 1$) electron liquids is
under good theoretical
 control. In this case, Boltzmann theory yields a good first approximation to the transport properties, and  any Fermi liquid corrections to the
bare electron mass and to the density of states are small.
Moreover, the screening radius $\lambda_{sc}=a_{B}^*/4$, is much
larger than the average distance between electrons: $\lambda_{sc}\
n^{1/2} = (4\sqrt{\pi})^{-1}r_{\text s}^{-1} \gg 1$.

 In this limit, to a first approximation, the transport scattering rate  can be calculated by perturbation theory.
Specifically, $\rho$ is proportional to the rate of transfer of momentum from the electrons to the lattice.
At high temperatures the resistance of the 2D electron gas is determined by the
electron-phonon scattering.
At  temperatures low compared to the effective Debye temperature, $T_{ph} \sim \hbar c k_{\text F}$, where $c$ is the speed of sound, the electron phonon scattering is no longer significant. (In some cases, a crossover temperature, $T_{ph}$, below which electron-phonon scattering is unimportant, can be readily identified; for example, in Figs. \ref{fig:fig2}, $T_{ph} > E_{\text F}$ is roughly the point at which $\rho(T)$ has a minimum.)

Since in all cases of interest here, the Fermi momentum is much smaller than the reciprocal
lattice vector, electron-electron scattering conserves the
total quasi-momentum to very high order, and therefore does not contribute directly to the
resistance. (This is different from metals with large Fermi momenta
where electron-electron Umklapp processes determine the $T$-dependence of the resistance
at low $T$ \cite{Abrikosov}.) Therefore the low temperature  resistance of the system  is due to electron-impurity scattering.

Consider, for example, the case in which the dominant low temperature scattering is from the
Coulomb potential due to a random distribution of charged impurities.
Naturally, the potential of an impurity is screened by the 2DEG, itself.  Two generic aspects of the screening are that it causes the potential to fall more rapidly when the distance from the impurity exceeds the screening length and it induces a rapidly oscillating component of the screened potential corresponding to the Friedel oscillations in the density.   As is well known, scattering by an unscreened Coulomb potential produces singular forward scattering, leading to an infinite cross section.  However, the transport scattering rate weights large angle scattering more heavily.  In 3D, the transport scattering cross section is still logarithmically divergent, so even when the screening length is long, it plays an essential role in cutting off this divergence.  In 2D, however, the transport scattering rate is finite, even in the limit $\lambda_{sc} \to \infty$, where in the Born approximation it is given by
\begin{equation}
\frac{\hbar}{\tau}=\frac
{(2\pi)^{2}e^4}  {E_{\text F}}
{N_{i}} \Big\{1 + {\cal O}\big([k_{\text F}\lambda_{sc}]^{-2}\big)\Big\}
\label{rate}
\end{equation}
where $N_i$ is the concentration of impurities.  Moreover, to the extent that quantum inference between different scattering events can be neglected (Boltzmann transport),
the conductivity is related to $\tau$, the Fermi velocity, $v_F$, and the density of states, $\nu$, according to the Drude formula
\be
\rho^{-1} = e^2 v_F^2 \tau \nu /2.
\ee
Here, $\rho$ approaches a constant value as $T \to 0$.
A  parallel magnetic field,  which we assume acts only on electron spins,
 changes the spin degeneracy at the Fermi level.
 In the limiting cases, $B_{\|}=0$, there is a spin degeneracy $g_{c}=2$, while for
$B_{\|}\gg B^{*}$,  $g_{c}=1$.  Since $E_{\text F} \propto v_F^2\propto   g_c^{-1}$, and $\nu\propto g_{c}$,
we conclude that for $r_{\text s}\ll 1$,
\begin{equation}
\rho\propto g_{c}.
\end{equation}
This means that the resistance {\it decreases} by  a factor of 2 as $B_{\|}$ increases from 0 to $B^\star$.
By the same token the scale for the $T$ dependence of $\rho(T)$ is set by $T_F=E_{\text F}/k_B$;  since at temperatures large compared to $T_F$, $\tau \propto v_F^2 \propto T$, it follows that $\rho$  {\it decreases} as $T$  increases in proportion to $T^{-2}$.
The only change in the $T$  dependence of $\rho$ in the presence
of $B_{\|}>B^{*}$ is produced by the factor of 2 increase in $T_F$.
Similar considerations (which we will not review explicitly) lead to the conclusion
that the drag resistance in double layers is small in proportion to $T^{2}/E_{\text F}$  and consequently $\rho_D$ decreases as $B_{\|}$ increases.

This simple theoretical description in the limit $r_{\text s}\ll 1$ and $k_{\text F}\ell \gg 1$ gives a good account of the transport in clean, high density devices, but is in drastic qualitative disagreement with  all the experimental results on the transport properties of the
large $r_{\text s}> 1$ devices presented in Section 1.

\subsection{Weak localization corrections and the theory of 2D localization}

In 2D, the  interference between multiple scattering processes, \emph{i.e.}, the ``weak localization corrections''   to the conductivity,  diverge at $T\rightarrow 0$, and $L\rightarrow \infty$ \cite{Abrahams,khmelnitski}. Here $L$ is the sample size. (For a review, see \textcite{Lee}.) This divergence, however, is only logarithmic, and therefore for $k_{\text F}\ell\gg 1$, and at accessible temperatures, these corrections are small in comparison to the zeroth order conductivity $G \equiv (h/e^2) \rho^{-1}= G_0+\delta G$, where $G_0 \sim k_{\text F} \ell$
    and
\begin{equation}
\delta G=-
\ln [L/\ell].
\label{weakloc}
\end{equation}
 In infinite samples $L$ in Eq.~\ref{weakloc} should be reinterpreted as
the phase breaking length $L_{\phi}=\sqrt{D\tau_{\phi}}$, where $D=v_F\ell$ is the diffusion constant and $1/\tau_{\phi} \sim T^{p}$ is the phase-breaking rate, in which $p$ depends on details of inelastic scattering processes.

\subsection{Interaction corrections}
\label{interactioncorrections}

Impurities in a metal create  Friedel oscillations of the electron
density. Due to the electron-electron interactions, the
quasiparticles in the
 metal are scattered
not only from the impurity but also from the modulations of the
 electron density.
The interference between
these two processes gives rise to
corrections to the Drude resistance.
 These corrections are interesting because they are non-analytic functions of $T$ and $B_{\|}$, so at small enough temperatures they dominate the $T$ and $B_{\|}$ dependences of the resistance.

In the diffusive limit ($L_{\text T}\equiv \sqrt{D/T}\gg \ell$)  the interaction correction to the conductivity is logarithmically divergent as $L_{\text T}\rightarrow \infty $  (\textcite{aal}; \textcite{Finkelshtein}; \textcite{Finkelshtein_1}; \textcite{Finkelshtein_2}; \textcite{CastellaniLee}; \textcite{AltshulerAronov}; \textcite{FinkelsteinMI1}):
\begin{equation}
\delta G=-\frac{1}
{2\pi^{2}
}\Big\{1+3\Big [1-\frac{\ln (1+F_{0})}{F_{0}}\Big ]\Big\} \ln (L_{\text T}/\ell),
\label{aronovaltshuler}
\end{equation}
where $F_{0}<0$ is an interaction constant in the triplet channel.
 Note that this same interaction parameter is also responsible for an enhancement of the spin susceptibility, $\chi =\chi_{0}/(1+F_{0})$. For $r_{\text s}\ll 1$ (in which limit, $|F_{0}|\ll 1$), Eq. \ref{aronovaltshuler} gives (up to a numerical factor) the same (negative) correction to the conductivity as Eq.~\ref{weakloc}. Note, however, that this correction has the opposite (``metallic'') sign when $\ln(1+F_{0})/F_{0}> 4/3$.

At higher temperatures, when $L_{\text T}\ll \ell$, the leading interaction effect involves the interference between a single  electron scattering from an impurity, and from the Friedel oscillations in the neighborhood of the same impurity. The interference corrections to the Drude formula in this so called ``ballistic regime''  were considered by \textcite{Aleiner,Aleiner1}.

 The result is that in an intermediate interval of temperatures, $E_{\text F}\gg T\gg  \hbar/\tau$, there is a $T$-dependent correction to the  conductance
which is is linear  in $T$. It consists of the sum of ``singlet" and ``triplet" contributions \cite{Aleiner,Aleiner1}:
 \begin{equation}
 G(T)-G(0)\propto  G_{0} \frac{T}{E_{\text F}} \Big [1 +3\frac{F_{0}}{1+F_{0}} \Big ].
 \label{CT}
 \end{equation}
 The factor of 3 in the second (triplet) part of Eq.~\ref{CT} reflects  the existence of 3 channels in the triplet part on electron-electron interaction. (Eq.~\ref{CT} is written for the case in which there is no valley degeneracy, as in GaAs, but it is readily generalized to the valley degenerate case. For example, in the case of Si MOSFETs, with a two-fold valley degeneracy, this factor becomes 15.) For repulsive interactions $F_{0}$ is negative with a magnitude which, for small $r_{\text s}$, is proportional to $r_{\text s}$.  Thus, at small $r_{\text s}$,  $\rho(T)$ decreases linearly as $T$ increases.

 Of course, most experimental realizations of the 2DEG have $r_{\text s} \gtrsim 1$.  At a phenomenological level, it is possible to imagine~\cite{Aleiner,Aleiner1} extrapolating to stronger interactions in the spirit of Fermi liquid theory, in which case for $F_0 < -1/4$, a linearly increasing resistance would result.  The range of validity of this sort of extrapolation, which is clearly sensible to some degree, is a unresolved theoretical issue in the field, to which we will return.

\subsection{The theory at $r_{\text s}\gg 1$}
\label{largerstheory}


In an ideal MOSFET,
where there is a metallic ground plane displaced by a
distance $d$ from the 2D electron gas, the electron potential energy changes its form as a function of $d$. For $r\ll d$ electrons interact via Coulomb interaction and
$V(r) \sim e^2/ r$. For $r \gg d$ one has to take into account the interaction between electrons and their images in the ground plane
and hence $V(r) \sim e^2d^2/ r^3$ .

\begin{figure*}
    \centering
         \includegraphics[width=.6\textwidth]{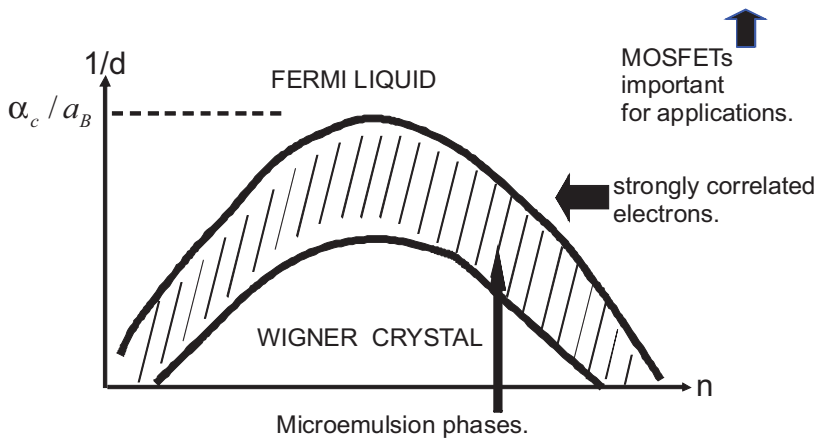}
        \caption{Schematic phase diagram for 2D electrons in a MOSFET with no disorder.}
    \label{fig:fig16}
\end{figure*}

The mean-field phase
diagram for this problem is shown schematically in Fig.\ref{fig:fig16}.
The considerations that produce the re-entrant character of transition line as a function of $n$ are relatively simple. The kinetic energy of the 2D electron liquid scale with the electron density as  $\sim n/m$. For Coulomb interactions, the typical interaction strength scales as $\sim e^2n^{1/2}$, so the ratio of the potential to kinetic energy, $r_{\text s}\propto n^{-1/2}$, decreases  with increasing density.  Therefore,  in the range of densities  $n \gg d^{-2}$, increasing density always favors the  FL.  Conversely, for $n \ll d^{-2}$, the interactions between electrons are dipolar, and hence effectively short-ranged, i.e. the typical interact scales as $\sim e^2d^2n^{3/2}$.  Thus, in this range of density, $r_{\text s}\propto n^{1/2}$  decreases with decreasing density, and hence decreasing density favors the FL, reflecting the same trends as does  $^3$He. One implication of this analysis is that there exists a critical value of the ratio, $d/a_B = \alpha_c$, such that for $d < \alpha_c a_B$, there is no WC phase at any density.
In this case there is a largest achievable value of ${\rm Max}[r_{\text s}] \sim d/a_B^\star$. 

Fixed node Monte Carlo simulations \cite{cip} performed under the {\it assumption} that there is a direct FL-WC transition yield the
 large critical value $r_{\text s}^{(c)}=38\gg 1$.
By analogy, we think that it is likely
that $\alpha_c\sim r_{\text s}^c$
 is also large compared to 1,
although  as far as we know, this issue has not been addressed using any
 quantitative methods.

 The very existence  of a ``highly correlated'' 2D 
 electron liquid 
 in the range $1 < r_{\text s} < r_{\text s}^{(c)}$ derives from the  large value of $r_{\text s}^{(c)}$.  From this point of view, it is possible to consider the case $r_{\text s} < r_{\text s}^{(c)}$, and still treat 
 $r_{\text s}$ as a large 
 parameter.  If we do this, we see that the highly correlated 2D 
 electron liquid 
 has
(at least) three characteristic energy scales:  1)  $E_{\text F}$, or more properly \cite{AndreevSQL,AndreevSQL1}, the
renormalized Fermi energy, $E_{\text F}^\star < E_{\text F}$, which contains
 a renormalized mass, 2) the interaction energy  $V=r_{\text s} E_{\text F}$, and 3) the plasma frequency
  $\Omega_P\sim \sqrt{E_{\text F} V} =\sqrt{r_{\text s}} E_{\text F}$. For large $r_{\text s}$, these energies are quite distinct,
   with $V > \Omega_P >E_{\text F}^\star$.
As a consequence of the existence of this hierarchy of energy scales
there are 4 distinct temperature intervals where the 
 electron liquid 
 behaves differently. We will discuss these intervals below:
$T<E_{\text F}$ where the system is in the Fermi liquid state,
$E_{\text F}<T<\Omega_{P}$ where the system is a non-degenerate strongly correlated and still highly quantum liquid, $\Omega_{P}<T<V$ where the system is a highly correlated classical liquid, and $T>V$ where the system is a classical electron gas.

\subsection{The Fermi fluid to Wigner crystal transition in the absence of disorder}

The classic studies of the Fermi fluid to WC transition
were carried out under the assumption that, at $T=0$,  there is a direct first order transition where the ground state energies cross.  However,
it has been shown (\textcite{italians}; \textcite{SpivakPS}; \textcite{SpivakKivelsonPS}; \textcite{Reza}; \textcite{SpivKivAnn}; \textcite{leib}; \textcite{chayes}) that for the case of a relatively long range potential
\begin{equation}
V\sim \frac{1}{r^{x}},    \,\,\,\,\,\, 1\geq x\geq 3,
\end{equation}
a direct first order liquid-crystal transition  is forbidden in 2D. Instead, either the freezing transition is continuous (which is unlikely, for well known reasons \cite{brazovski}), or between these two phases there must occur other phases, which we have called ``microemulsion phases" .

The largest theoretical uncertainty, here, is that no reliable estimates exist concerning the width in density of these novel phases.  The part of the phase diagram where these phases should exist is indicated qualitatively by the shaded region in Fig.\ref{fig:fig16}.  A rich variety of phases, including bubble and stripe phases, are expected in this region.  The existence of a stripe phase, for example, would be detectable macroscopically through a spontaneously generated resistance anisotropy.

 At the moment there is no direct experimental evidence of the existence of such phases. This is either due to the fact that, for some reason, the range of densities in which they exist is extremely small, or because
 currently available samples are not pure enough to reflect the physics of the zero disorder limit.  In any case, the
 existence of such phases, which are very different from both the FL and the WC, significantly complicates the theory of
strongly correlated disordered electronic systems.

\subsection{Theoretical considerations concerning the metal-insulator transition in 2D with disorder}

\subsubsection{Anderson localization}
It has been suggested \cite{Abrahams} that in absence of electron-electron interactions the logarithmic derivative
\begin{equation}
\frac{d \ln G}{d \ln L}=\beta(G)
\label{weaklocalization}
\end{equation}
is a function of $G$ alone.
According to Eq.~\ref{weakloc},
$\beta\sim -1/G$ as $G\to \infty$.
The fundamental result of the single particle theory of localization 
follows from this:  even weak disorder leads to insulating behavior at $T=0$, and hence there can be no metal-insulator transition.
On the face of it, this statement is inconsistent with the experiments summarized in the previous section.

\subsubsection{Transition in the presence of spin orbit scattering}

Since the existence of a metal insulator transition is an important issue, it is worth considering the single particle problem in the presence of spin orbit coupling, where the theory predicts a zero temperature transition between an ``ideal metal" ($G=\infty$) and an insulator ($G=0$) (\textcite{Larkin}; \textcite{tando}; \textcite{evangelou}). Analytically, this follows again from a perturbative analysis of the interference corrections:  $\delta G = +\frac{1}{4} \ln(L/\ell_{so})$ for  $L$ large compared to the spin-orbit scattering length,  $L\gg \ell_{so}\geq \ell$.  In other words, so long as $G$  at length scale $ \sim \ell_{so}$ is large compared to 1, at low enough temperatures that $L_{\phi} > \ell_{so}$, the conductivity is an {\it increasing} function of decreasing $T$ -- ``antilocalization.''
Under the same assumption of one-parameter scaling as in Eq. \ref{weaklocalization}, this implies that $\beta \sim + 1/G$ as $G\to \infty$, which in turn implies the stability of a perfectly  metallic phase with $\rho=0$.
On the other hand, the stability of an insulating phase for strong enough disorder is not debatable, {\it i.e.} $\beta < 0$ for small enough $G$.
Moreover, the validity of the one-parameter scaling in this problem and the existence of a metal insulator transition have been tested in numerical studies \cite{Asada}.

While this is the simplest model system in which a 2D metal insulator transition exists, it is unlikely that it is relevant to the experiments discussed in Section 1.
Specifically, the spin orbit interaction in n-silicon MOSFETs is quite weak, and hence unlikely to be important in the accessible range of temperatures.

\subsubsection{Scaling theories of a metal-insulator transition with strong interactions}

Even in the absence of spin-orbit coupling, assuming the theory remains renormalizable, the $\beta$ function is fundamentally modified by strong interactions. A key idea underlying most scaling theories of the metal-insulator transition (\textcite{McMillan}; \textcite{FinkelsteinMI1}; \textcite{sudipandelihu}; \textcite{FinkelsteinMI2}; \textcite{FinkelsteinMI3}) is that the logarithmically diverging interaction correction to the conductivity in Eq. \ref{aronovaltshuler} can have the opposite sign and larger magnitude than the single particle weak localization corrections in Eq.~\ref{weakloc}  when
 extrapolated to large values of $r_{\text s}$ where $|F_{0}|$ is no longer small.
 Moreover, even where perturbation theory is still valid, {\it i.e.}, where $G\gg 1$ and $r_{\text s} \ll 1$,
 the value of $F_{0}$
 is subject to  renormalization, which in the case of
 weak Coulomb interactions,
 $|F_{0}|\ll 1$, is of the form \cite{FinkelsteinMI1}
 \begin{equation}
 \frac{d F_{0}}{d \ln (L/\ell)}\sim -\frac{1}{k_{\text F}\ell}.
 \end{equation}
 Since  $F_{0}$ is negative, this means that its magnitude grows with
 increasing $L$.
 It is therefore possible to imagine
  (in the spirit of the renormalization group, RG) that even where the bare interactions are weak, at sufficiently long scales, $F_0$ grows until
 $|F_{0}|\lesssim 1$, at which point
 the $\beta$ function would change sign.

If a metal insulator transition occurs, it  must occur at finite (dimensionless) interaction strength, $F^\star$, and finite conductance, $G^\star$.  (Here, $F$ may denote the strength of one or several interactions.)  \emph{A priori}, the properties of such a fixed point cannot be computed perturbatively.  Either, the existence and character of such a fixed point must be inferred by extrapolating perturbative expressions for the $\beta$ function to finite coupling strength 
\cite{FinkelsteinMI1,CastellaniLee,FinkelsteinMI2,FinkelsteinMI3,Chamon,Nayak}, or it must simply be conjectured \cite{sudipandelihu} on phenomenological grounds.
One problem with this scenario is that for $G\gg 1$, if the RG procedure is taken literally, the
 system evolves to a low temperature state where $F_{0}\rightarrow -1$,
which in addition to changing the sign of the $\beta$ function, implies the existence of a magnetic instability, which surely must affect the physics.
More generally, there exists no clear qualitative picture of what happens to the system when
the parameter $F_{0}$
 is significantly renormalized.

There are a number of experimentally relevant consequences of this scenario.  Most importantly, it implies the existence of a true quantum phase transition between a (possibly ideal) metal  and an insulator, with all the implied quantum critical phenomena.
Secondly,  valley degeneracy figures as an important factor for the character, and possibly even the existence of a metal-insulator transition \cite{FinkelsteinMI2,FinkelsteinMI3}. Thirdly, the theory predicts the existence of a peak in the temperature dependence of $\rho(T)$. For systems with $k_{\text F} \ell \gg 1$, $\rho(T)$ reaches a maximum as a function of $T$ at the temperature \cite{FinkelsteinMI1,FinkelsteinMI2,FinkelsteinMI3}
 \be
 T_{\text {max}}\approx (v_{F}/\ell)\exp(-\alpha 
 G)
 \label{Tmax}
 \ee
  corresponding to the length scale, $L(T_{\text {max}})$, at which $\beta$ changes sign. Here $\alpha$ is a constant of order one, which depends on the strength of the bare interactions.  For $T < T_{\text {max}}$, the system exhibits ``metallic'' $T$ dependence. The theory also predicts an increase of the resistance at small values of $B_{\|}$.  However  when the conductance of the system is large, $G \gg 1$, both these effects become very small
and manifest themselves only at exponentially low temperatures (see Eq. \ref{Tmax}).

\subsection{2DEG in the presence of a perpendicular magnetic field}
\label{PerpMagnFtheory}

In considering the phase diagram of the 2DEG in the $n$ - $B_{\perp}$ plane with a fixed strength of disorder, there are a number of asymptotic statements \cite{global} that can be made with confidence on theoretical grounds:
\begin{itemize}
\item Firstly, for low enough density $n < n_{min}$ and for any strength of magnetic field the system must be insulating;  in the absence of disorder, this is a consequence of Wigner crystallization, and in the presence of disorder, it can be traced to the strong tendency of dilute electrons to be strongly localized by disorder.
\item  For any fixed density, and high enough magnetic field, $n\phi_0/B_{\perp} \ll 1$, it is similarly straightforward to show that the system must always be insulating. (Here, $\phi_0=hc/e$ is the quantum of flux.)
\item Most interestingly, in the limit that both $n$ and $B_{\perp}$ are large, but with $n\phi_0/ B_{\perp} \sim 1$, there necessarily exist robust quantum Hall phases.  To see this, note that in this limit, $\hbar\omega_c$  is large both compared to the strength of the disorder and the electron-electron interactions, so inter-Landau level scattering can be treated perturbatively.  (Here, $\omega_c=eB_{\perp}/mc$.)  In this limit, the existence and stability of integer and fractional quantum Hall states is well established.
\end{itemize}

For present purposes, we will neglect the interesting complexity associated with the various distinct quantum Hall phases, and simply discuss the considerations that determine the shape and topology of the curve, $n^\star(B_{\perp})$, that  encloses the regions of quantum Hall phases in the phase diagram (see Figs.\ref{fig:fig8} and \ref{fig:fig9}).  From the three observations above, it follows that
\begin{equation}
n^\star(B_\perp) \sim B_{\perp}/(\phi_0) \ \ {\rm as}\ \ B_\perp \to \infty .
\end{equation}
 In the absence of electron-electron interactions and spin-orbit scattering, we know that all states at $B_{\perp}=0$ are localized.  This requires that $n^\star(B_{\perp})\to \infty$ as $B_{\perp}\to 0$.  Within the single-particle theory, this constraint was accounted for by the notion of ``levitation'' of the delocalized states \cite{khmelnitskii84,laughlin84}.  The idea here is that in the large field limit, when $\omega_c \gg 1/\tau$,
 there is a single delocalized energy level at the center of each Landau level, but that when at small field, $\omega_c \sim 1/\tau$, the delocalized states initially associated with each Landau level remain distinct, but begin to ``levitate'' to higher energies.  Thus, as $B_{\perp}\to 0$, the energies of the delocalized states diverge.  More recent numerical studies \cite{sheng-2001} have suggested that, even for non-interacting electrons, the fate of the delocalized states may be a more complex issue. However, clearly, so long as there is only a single, insulating phase at $B_{\perp}=0$, the divergence of $n^\star(B_{\perp})$ as $B_{\perp}\to 0$ is inescapable.

Conversely, if $n^\star(B_\perp)$ tends to a finite value, $n_{\text c}=n^\star(0)$, as $B_\perp \to 0$, it probably implies the existence of a zero field phase transition at this value of the density.  Since $n^\star$ is unambiguously a critical density marking the point of a quantum phase transition, tracking it to the zero field limit is a very promising strategy for distinguishing a crossover from a true phase transition.

Comparing these theoretical expectations with the already discussed experiments, we see that in the large $r_{\text s}$ p-GaAs and Si MOSFETs (the right hand side panel in Fig.\ref{fig:fig7} and Fig.\ref{fig:fig9}), $n^\star(B_\perp)$ clearly extrapolates to a finite value as $B_\perp \to 0$.  Conversely, in the small $r_{\text s}$ n-GaAs device (left hand side panel in Fig.\ref{fig:fig7} and Fig.\ref{fig:fig8}), clear evidence of levitation is seen, in the sense that $n^\star(B_\perp)$ has a pronounced minimum at a value of $B_\perp$ which arguably corresponds to $\omega_c\tau \sim 1$, and then grows strongly as $B_\perp$ is further reduced.

\subsection{Microemulsions of WC and FL in the presence of weak disorder}

\subsubsection{Effects of weak disorder at $T=0$}

Starting from the clean limit, treating the interactions as strong and the disorder as a small perturbation leads to a rather different perspective on the metal insulator transition.   It is important to recognize, however, that  even weak disorder is always  a relevant perturbation in 2D.  If we ignore the existence of microemulsion phases in the zero disorder limit, then in 2D disorder always rounds the first order liquid to WC transition (\textcite{Imry}; \textcite{Aizenman}; \textcite{Baker}), resulting in a phase in which islands and continents of pinned WC coexist with seas and rivers of Fermi fluid.  If, instead, we start with a spatially ordered pattern of alternating WC and Fermi fluid, characteristic of a microemulsion phase, then disorder destroys the long-range order, again resulting in a phase which consists of a disordered mixture of WC and Fermi fluid regions.  Though the two scenarios differ greatly for weak disorder in the degree of local organization of the coexisting regions, at long distances they are difficult to distinguish.

 Since at $T=0$, a WC does not conduct, this picture calls to mind a percolation type
  metal-insulator
  transition.  Of course, this picture
  does not take into account a variety of possibly important effects, including
interference effects
 which could ultimately turn the percolative metal into an insulator.  However,  in the limit
 in which  the characteristic sizes of the islands and continents are large, classical percolation could be a good description over a significant
 range of $T$.
 
 \begin{figure*}
    \centering
         \includegraphics[width=.5\textwidth]{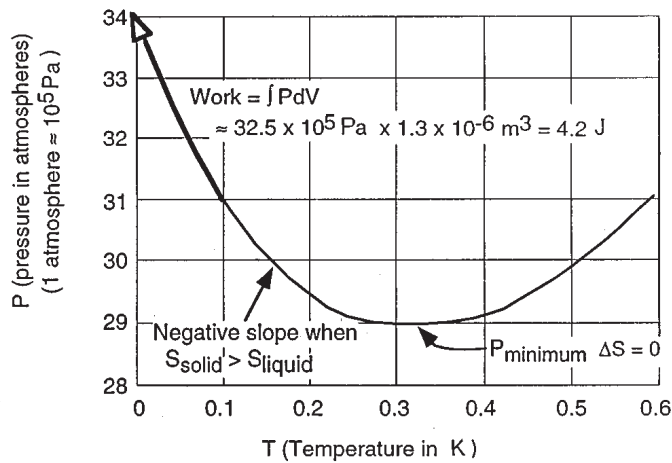}
        \caption{The melting pressure of $^3$He. Adapted from \textcite{richardson97}.}
    \label{fig:fig17}
\end{figure*}

\subsubsection{Low $T$ thermal physics:  the Pomeranchuk effect}
\label{Pomeranchuk}

The Fermi-liquid to crystal transition takes place at $r_{\text{s}}\gg 1$.
In the Fermi liquid state,  the entropy density, $S_{\text{FL}}\approx
n\frac{T}{E_{F}}$, is small
for temperatures   $T \ll E_F$.
In contrast,
for $T$ large compared to
$J$, the spin exchange interaction between localized particles in the
crystal state,
 the spin entropy of the crystal
$S_{\text{C}} \approx n \ln2 $
  is relatively large and temperature independent.
Since for large $r_{\text s}$, $J$ is exponentially
smaller than $E_F$
 (See, for example,  \textcite{Roger}.),
 there exists a broad range of temperatures, $E_{F}>T>J$, in which the crystal
is the high-entropy state, $S_{\text{C}}  >
S_{\text{FL}}$.
In this range of temperatures and near the point of the crystal-liquid
transition  the system tends to freeze upon heating!
  Of course,
  at ultra-low temperatures,  $T\ll J$, or at high temperatures, $T>E_{F}$
   the entropy of the crystal
   is smaller than that of the liquid, so the Pomeranchuk effect disappears.

The tendency of Fermi systems to freeze upon heating, known as the ``Pomeranchuk
effect,''  was originally
discovered in the framework of the theory of $^3$He  \cite{Pomeranchuk,richardson97}, where
it has been confirmed experimentally. (See the phase diagram of $^3$He in
Fig.\ref{fig:fig17})  In this regard, the Wigner crystal of electrons and the crystal of
$^3$He atoms are similar.  Specifically, the exchange energy between spins in a
Wigner crystal is
exponentially small, $J \sim \exp[-\alpha \sqrt{r_s}]$ where $\alpha$ is a
number of order one (\textcite{Roger}; \textcite{sudip1}; \textcite{sudip2}).
 Thus at low temperatures the  evolution of the random microemulsion phases as a
function of $T$ and $B_{\|}$ is dominated by the entropy changes associated with
the spin degrees of freedom
\cite{SpivKivAnn,SpivakPS,SpivakKivelsonPS}. 

In the case of the two phase coexistence at low temperatures,
  the fraction of the system which is locally Wigner crystalline, $f_{\text {WC}}$, increases linearly with $T$. 
  This behavior will dominate the $T$ dependence of many important physical properties of the system at low $T$.

 Similar considerations govern the $B_{\|}$ dependence  of the phase diagram.
Since the spin susceptibility $\chi_{\text {WC}}\gg \chi_{\text {FL}}$, the corresponding magnetization $M_{\text {WC}}\gg M_{\text {FL}}$ at small $B_{\|}$.  Since the free energy of the system contains the term $-M B_{\|}$, there is a $B_{\|}$-induced increase of the WC fraction over a wide range of circumstances.

 Since the resistance of the WC and FL are very different, the  Pomeranchuk effect
is one of the most directly testable features of the microemulsion phases.   Where the FL is the majority phase, it leads to the robust prediction that  $\rho(T, B_{\|})$ increases with $T$ and $B_{\|}$ at low $T$,
reflecting the purely thermodynamic  fact that $f_{\text {WC}}$  increases as $T$ and $B_{\|}$ increase.
However, for $B_{\|}>B^{\star}$,  the $T$ and $B_{\|}$ dependences of $\rho(B_{\|})$ are  quenched since there is no spin-entropy in the fully polarized system, so $f_{\text {WC}}(T, B_{\|})$
no longer depends strongly on these quantities.
This effect can, in principle, produce arbitrarily large fractional changes in the low $T$ resistivity.  Even on the insulating side of the transition, where at $T=0$ a majority of the system is Wigner crystalline, a temperature induced increase in $f_{\text {WC}}$ can result in an increase of $\rho$ with increasing $T$, in a limited range of temperatures, even though $\rho > h/e^2$.

Many other, more detailed aspects of the $T$ and $B_{\|}$ dependences of $\rho$ depend on  the type of disorder and other ``details.''  While some progress has been made in understanding the dynamical properties of microemulsion phases in the presence of weak disorder \cite{SpivKivAnn}, no satisfactory and/or quantitative theory of disordered microemulsions is currently available.

 Of course, at high enough temperatures, 
 where the electron liquid is non-degenerate, 
 the WC always is the lower entropy phase.  
The Pomeranchuk effect reflects, more than anything, the low entropy of the degenerate Fermi liquid at $T < E_{\text F}$.  
 The melting temperature of the WC is determined by the (possibly strongly renormalized) Fermi energy of the ``competing'' fluid.  Above this temperature, the microemulsions give way to a uniform non-degenerate electron fluid.

\subsubsection{Crossovers at higher temperatures ($T>T_{\text F}$) and large $r_{\text s}$}
\label{crossoverhighT}
In the absence of disorder, Umklapp scattering, and electron-phonon scattering, the long-wave-length properties of the electron fluid are governed by hydrodynamics, rather than by the Boltzmann equation.  In this limit, we need to talk about the viscosity of the fluid, $\eta$, rather than the conductivity, which is in any case infinite.  For gentle enough disorder, we can think of the disorder as defining some form of effective medium through which the otherwise hydrodynamic fluid flows.  This is a transport regime, which is not often recognized, in which the hydrodynamic healing length ({\it e.g.}, the electron-electron mean-free path, $\ell_{\text{e-e}}$) is short compared to the distance between impurities, or the length scale over which the disorder potential varies.  In this case,
\be
\rho \propto \eta
\label{rhoeta}
\ee
where the proportionality constant is in general complicated function of the strength and character of the disorder potential, and $\eta$ is the viscosity of the electron fluid in the absence of disorder.  The conditions for the  validity of this equation  are generally violated at $T\to 0$, where $\ell_{\text{e-e}} \to \infty$, but for sufficiently strong interactions, it can be satisfied down to moderately low $T$.

Here, we make a few remarks concerning the viscosity of the uniform fluid at $T>T_{\text F}$.  As mentioned previously, for large $r_{\text s}$, there is a broad range of temperatures in which $T_{\text M} \sim T_{\text F}  < T < V$, where the electron fluid, although non-degenerate, is still highly correlated. (Here $T_{\text {M}}$ is the melting temperature of the Wigner crystal.) It is generally observed that the viscosity of highly correlated fluids is a decreasing function of increasing temperature. This observation, combined with Eq. \ref{rhoeta} and the low temperature increase in resistivity produced by the Pomeranchuk effect,  implies that for metallic samples with large $r_{\text s}$, there should generally be a maximum in the resistivity at $T \sim T_{\text F}$.

Looking at this problem more carefully, there are, as noted in Section \ref{largerstheory}, two distinct ranges of $T$ to be considered.  Beyond this, our theoretical understanding of the viscosity of correlated fluids is crude at best. We thus rely on the following line of arguments to get a feeling for the expected $T$ dependence of $\eta$. (i) For $V>T>\Omega_{\text {P}}, T_{\text {F}}$ the electron fluid is a highly correlated classical fluid.  There are many examples of such fluids - indeed, most classical fluids fall in this regime \cite{Frenkel}. The viscosity of classical fluids is widely observed (\textcite{Frenkel}; \textcite{DAD}) to be an exponentially increasing function of decreasing temperature.
(ii) For $T_{\text F}\ll T \ll \Omega_{\text {P}}=\sqrt{E_{\text F} V}$  the fluid is still quantum but not degenerate, and is still strongly correlated.
 It has been conjectured on theoretical grounds \cite{AndreevSQL1} (see also \textcite{SpivKivAnn}) that in this regime
\begin{equation}
\eta(T) \propto  1/T.
\label{voscosity}
\end{equation}

\section{Theoretical interpretations of experiment}

We now discuss some of the attempts that have been made to interpret the corpus of experimental observation summarized in Section 1 in terms of the various theoretical results outlined in Section 2.  As is clear from Section 2, there is presently no well controlled theory that treats non-peturbatively both the  strong correlation  and the disorder effects, so all such attempts involve an extrapolation of results from small $r_{\text s}$ to large $r_{\text s}$, from zero disorder to finite disorder, or both. Thus, while we present arguments both in support of and against various proposed interpretations, none of our conclusions are irrefutable.

\subsection{Explanations based on classical ``Drude'' formulas}
\label{classicalexplanation}

A systematic attempt to explain a wide range of the experiments presented in Section 1 has been undertaken by \textcite{Sternsarma},  \textcite{dolgopolov1}, \textcite{dolgopolov2}, and \textcite{SarmaHwang,SarmaHwang1,SarmaHwang2,SarmaHwang3}  employing the classical  formulas for the resistivity which are valid for weak scattering and $r_{\text s}\ll 1$ and extrapolating them to the case $r_{\text s} \gg 1$.  Manifestly, this approach involves extrapolating results from the weak interaction regime into a regime in which they are strong.  Since the  electron impurity scattering is treated perturbatively (Born approximation) and the electron screening is computed at the level of RPA,  there is no formal justification for the approach when $r_{\text s} > 1$.  However, the appeal of this approach is that it leads to explicit expressions for a wide range of physical quantities which can be directly compared with experiment.  A number of  striking quantitatively successful comparisons between this theory and experiment have been reported by \textcite{SarmaHwang,SarmaHwang1,SarmaHwang2,SarmaHwang3}.

 There are, however, several aspects of this extrapolation that we find troubling.

The extrapolation to large $r_{\text s}$  does not simply involve quantitative shifts, but qualitative changes.   Whereas at small $r_{\text s}$,   as discussed in Section \ref{classicaltheory}, both $d \rho/d T <0$ and $d \rho / d B_{\|} < 0$,  the classical formulas extrapolated to large $r_{\text s}$ exhibit the opposite sign trends, $d \rho/d T >0$ and $d \rho / dB_{\|} >0$.  The change of sign at $r_{\text s}\sim 1$ does, admittedly, bring the results into qualitative agreement with experiment on large $r_{\text s}$ systems, but it cannot be said to be a ``featureless'' extrapolation. Moreover, the origin of the sign change can be traced to the fact, already discussed above, that at large $r_{\text s}$, the screening length $\lambda_{\text {sc}}$ obtained in RPA approximation is parametrically smaller than the spacing between electrons, $\lambda_{\text {sc}}\sqrt{\pi n} = (1/4) r_{\text s}^{-1}     \ll 1$.  Screening lengths less than the distance between electrons are clearly unphysical, and we worry that the same is true of other extrapolated results.

This same approach has been extended to explain the decrease of  $\rho(T)$ as a function of increasing $T$ at $T>T_{\text {max}}\sim T_{\text F}$, where the electron gas is non-degenerate.  Here, the resistivity is computed using the kinetic theory of a weakly interacting electron gas scattering from Coulomb impurities \cite{SarmaHwang}.  In the framework of this approach the decrease of the resistance is associated with the increase of the velocity and decrease of the scattering rate as $T$ increases. As a result $\rho \propto 1/T$. Again, semiquantitative success has been reported in the comparison of this theory with experiment. Since this theory neglects all correlations in the electron gas or between scattering events, it is well justified when $G \gg 1$ and  $r_{\text s} \ll 1$  for all $T > T_{\text F}$, and even for $r_{\text s} \gg 1$ at temperatures $T \gg V$.  However, the experiments in question  deal with the temperature range $T_{\text F} < T < V \sim r_{\text s} T_{\text F}$ in
samples with $r_{\text s}\gg 1$;  in this regime, the electron liquid is still strongly correlated, although not quantum mechanically coherent.

Additionally, Drude theory does not incorporate an insulating state, much less a metal-insulator transition.  It seems clear to us that the anomalous behavior of the metallic state at large $r_{\text s}$ are related to the appearance of a  metal-insulator transition in the same devices.

 \subsection{Interaction corrections to the conductivity in the ``ballistic
 regime'' $L_{\text T}\ll l$.}

 For $r_{\text s} \ll 1$ and at high enough temperatures that $L_{\text T} \ll \ell$ but low enough that $T \ll T_{\text F}$, the contribution ~\cite{Aleiner,Aleiner1} to the resistivity of electrons scattering from the Friedel oscillations induced by impurities is given by Eq. \ref{CT}. At small $r_{\text s}$, where $| F_0 | \ll 1$,  this expression gives the opposite sign of $d\rho/dT$ than is seen in experiments on samples with $r_{\text s} \gg 1$ and $G\gg 1$.  However, when Eq. \ref{CT}  is extrapolated to large enough $r_{\text s}$, where plausibly $F_{0}<-1/4$, $d\rho/dT$ changes sign,  and $\rho(T)$ becomes a linearly increasing function of $T$. (Where there is valley degeneracy, as in the case of Si MOSFET's where there are two valleys, the same sign reversal occurs  when $F_{0}<-1/15$ \cite{Aleiner,Aleiner1}.)  The same theory \cite{Aleiner,Aleiner1} predicts a decrease in the magnitude of the triplet part of Eq.~\ref{CT}   by a factor of $3$  when the electron  gas is spin polarized by the application of  $B_{\|}>B^{*}$. The reason is that the singlet and $L_{\text {z}}=0$ triplet parts of the two particle propagator are unaffected by $B_{\|}$, while the remaining two components of the triplet are suppressed by $B_{\|}$.

While this approach also involves an extrapolation, involving a sign change at $r_{\text s} \sim 1$, there is certainly nothing unphysical about a Fermi liquid parameter with a substantial magnitude:  $F < -1/4$. However, there are other aspects of this explanation of the experiments  presented in Section 1 that we find problematic.

 Firstly, this is a theory of {\it corrections} to the conductivity; even when the result is extrapolated to $r_{\text s} \sim 1$ and $T\sim T_{\text F}$, this approach only makes sense if the correction is small compared to the Drude conductivity itself, $G_{0}$.  Since the correction is a non-analytic function of $T$, it can be the dominant contribution to $d\rho/dT$ for sufficiently small $T\ll E_{\text F}$.  However, it is something of a stretch to interpret the large fractional changes in the conductivity seen in experiments on samples with $r_{\text s} \gg 1$ and $G\gg 1$ in these terms.  Over a range of temperatures and fields, $T\sim E_{\text F}$ and/or $B_{\|}\sim   B^*$, the analytic variations  of the Drude conductivity (which were the focus of the theory discussed in Section \ref{classicalexplanation}) should generally be at least as large as these corrections.
 In more physical terms, it is hard to imagine that the scattering of the electrons from the induced Friedel oscillations of the electron density  can make a larger contribution to the scattering cross-section than the scattering from the impurity itself. Therefore, we think that this process (even when extrapolated to large $r_{\text s}$, and taking the most optimistic viewpoint)    cannot explain changes in the resistivity by more than a factor of two.  However, in experiment, $\rho$ is observed to increase (admittedly more or less linearly with $T$) by a factor of 4 in large $r_{\text s}$  p-GaAs heterostructures, and by a factor of 10 in Si MOSFETs.

Secondly, although this theory does predict a suppression of the $T$ dependence of the resistivity by a parallel magnetic field, $d \rho(T)/d T|_{B_{\|}=B^{*}} \ /\ d \rho(T)/d T|_{B_{\|}=0} < 1$, the measured ratio at low $T$ is significantly smaller than that predicted by the theory  \cite{Aleiner,Aleiner1}.

In short,  we conclude that, although the results of the perturbation theory (\textcite{Sternsarma}; \textcite{dolgopolov1}; \textcite{dolgopolov2}; \textcite{Aleiner,Aleiner1}) are likely relevant to experiments performed on samples with relatively small $r_{\text s}$ and at relatively small values of $B_{\|}$ and $T$, they cannot explain results of experiments at $r_{\text s}\gg 1$ where almost all effects are of order one or larger.

\subsection{Scaling theories of the metal-insulator transition in 2D}

The scaling theories discussed in the previous section imply the existence of a metal-insulator transition at a sample dependent critical density, $n=n_{\text c}$.  At criticality, the resistance  approaches a finite limit, $\rho_{\text c}$, in the $T\to 0$ limit which is probably universal.  Strong $n-n_{\text c}$ and $T$ dependences of the resistivity near criticality are governed by universal critical exponents and scaling functions.  The appeal of such an approach is the robustness implied by universality. However, our principle interest is with the behavior of samples over a broad range of $r_{\text s}$. Thus, other than to stress the importance of the existence of the transition itself, we have chosen not to focus particularly on experiments very close to criticality. On the basis of the data in Fig. \ref{fig:fig1}, we identify an empirical value for the critical resistance, $\rho_{\text c} \approx 1 (h/e^2)$.  Thus, we can identify samples with low temperature resistances $\rho > \rho_{\text c}$ as ``insulating'' and with $\rho < \rho_{\text c}$ as ``metallic''.

Any theory which is based on electron interference effects, in practice predicts substantial fractional changes of $\rho(T,B_{\|})$ over plausibly accessible ranges of $T$ only in the near vicinity of the critical point.  In particular, where $G \gg 1$, only fractionally small logarithmic variations are predicted.  In contrast, in all systems which we describe in this review, the transport anomalies take place in a wide range of the  electron densities: up to five times the critical density (see, \textit{e.g.}, \textcite{cond-mat/0008005}). Therefore, most of the experimental data do not lie in the critical region. The experiments deep in the metallic region $\rho(T,B_{\|})$ exhibit very large changes in the resistance (sometimes by the order of magnitude) even in samples where the conductance is as large as $G\sim 10-20\,e^2/h$ (see Figs.\ref{fig:fig1}-\ref{fig:fig4}). In fact, the temperature dependence of the resistivity significantly increases in samples which are farther away for the critical region.

We therefore conclude that, although this theory quantitatively describes experimental data on Si MOSFETs in the close vicinity of the transition \cite{anissimova07}, it cannot explain large effects far from the transition.

\subsection{Interpretations based on electronic microemulsions and the Pomeranchuk effect}

In an electron fluid with  $r_{\text s} \gg 1$, the interaction energy is large compared to the kinetic energy, and so the short range correlations must certainly be  Wigner crystalline like, whatever the long-range emergent properties.  No theoretically well controlled treatment of this problem, capable of quantitative comparison with experiment, currently exists that treats on an equal footing the short-range crystalline correlations and the long-distance fluid character of the state.  The nascent theory of electronic microemulsions is, however, an attempted first step in this direction.  Here, we sketch the ways that many of the most significant qualitative aspects of the experiments can be understood from this perspective.

\subsubsection{Interpretation of the $T$ and $B_{\|}$ dependences of the resistance for $T < T_{\text F}$}

The  $T$ and $B_{\|}$ dependences of the resistance at $T<  T_{\text F}$ can be qualitatively understood as consequences of  the Pomeranchuk effect.  As the fraction, $f_{\text {WC}}$, of Wigner crystal grows with increasing $T$ and $B_{\|}$, this naturally produces an increasing $\rho$, {\it i.e.} a ``metallic'' $T$ dependence and a positive  magneto-resistance.  Moreover, to the extent that the exchange energy, $J \ll k_{\text B}T$,  $f_{\text {WC}}$ is a strongly non-analytic function which, for $B_{\|}=0$ and $T \ll T_{\text F}$ is linear in $T$,  and for $T\ll B_{\|}$ is an initially  linearly increasing function of  $B_{\|}$ which  saturates at $B_{\|}>B^{*}$.  Moreover, the $T$ dependence of $f_{\text {WC}}$ is quenched by $B_{\|}$ in the range $T < B_{\|}$.

As seen in Section 1, the experimentally measured $\rho$ in samples with $G \gg 1$ and  $r_{\text s} \gg 1$ exhibits all the same qualitative dependences as $f_{\text {WC}}$, including the dramatic 1 - 2 order of magnitude decrease  in $d\rho/dT$ produced by a field $B_{\|}> B^{*}$.  The transport theory of microemulsions \cite{SpivKivAnn} which relates $\rho$ to $f_{\text {WC}}$ is complex and incomplete.  In the limit that the disorder potential is smooth and $f_{\text {WC}} \ll 1$, the dominant contribution to the resistance comes from electrons scattering from rare islands of WC, which are themselves pinned at minima of the disorder potential.  In this case, it is easy to see that $\rho \propto f_{\text {WC}}$.

For other forms of disorder, and in all cases  where $f_{\text {WC}}$ is not small, the dependence of  $\rho$ on $f_{\text {WC}}$
is  more complicated, although, presumably, $ \rho$ is still a monotonically increasing function of $f_{\text {WC}}$. Thus,
the Pomeranchuk effect gives a plausible explanation of the giant positive magneto-resistance in the ``insulating'' regime, $\rho > h/e^2$, of the sort shown in Fig.\ref{fig:fig12}. It also gives rise to  the possibility that samples with $\rho>h/e^{2}$ can still have a metallic temperature dependence $d \rho/d T>0$.  Such behavior is occasionally seen, as shown in Fig.\ref{fig:fig11}, although so far only relatively small effects in samples with $ \rho$ no larger than $ \sim 3 h/e^2$. However, in principle, the Pomeranchuk effect can produce arbitrary large effects. In particular, on a qualitative level it can explain the existence of a metal-insulator transition as a function of $B_{\|}$.

\subsubsection{Interpretation of the magnteto-resistance in a perpendicular magnetic field}

Moving to the effect of a perpendicular magnetic field, the experimentally observed behavior can be quite complex, as shown in Figs.\ref{fig:fig6}, \ref{fig:fig13} and \ref{fig:fig14}. As is confirmed in tilted field experiments (also shown in the Fig.\ref{fig:fig14}), this complexity reflects a combination of spin and orbital effects.  The strong negative magneto-resistance seen at large fields, where the spins are fully polarized, generally reflects the existence of quantum Hall phases, as discussed in Section \ref{PerpMagnFtheory}.  For intermediate values of $B_\perp$, where spin physics is important, the Pomeranchuk effect can readily account for the existence of a large positive magneto-resistance, as discussed above. (The small negative magneto-resistance, sometimes seen in metallic samples at small values of $B_{\perp}$ (see Fig.\ref{fig:fig6}), is presumably due to weak localization effects.) This general analysis applies either to samples in which the zero field resistivity is less than $h/e^2$  (Fig.\ref{fig:fig6}) or greater than $h/e^2$ (Figs.\ref{fig:fig13} and \ref{fig:fig14}). In the more resistive samples, a part of the negative  magneto-resistance is presumably due to interference corrections to variable range hopping \cite{ShklovskiiSpivak,ShklovskiiSpivak1}.

\subsubsection{Interpretation of the $T$-dependence of the resistance at $T>T_{\text F}$ }

The existence of a broad range of temperatures with $T_{\text F} r_{\text s} \sim
V > T > T_{\text F}$ in which the  electron fluid is still strongly
correlated is one of the most clearly inescapable, and at the same
time widely neglected features of the discussion of the 2DEG with
large $r_{\text s}$.  The best understanding we have of the transport
theory, here, is obtained by arguing in analogy with other
strongly correlated fluids. As discussed in Section
\ref{crossoverhighT}, because there is no coherence, the Fermi
statistics of the electrons is relatively unimportant, but so long
as $T_{\text F} \sqrt{r_{\text s}} > T > T_{\text F}$, quantum effects could be crucial.
Moreover, because the interactions are strong and the samples are
``clean,'' it is plausible  that $\rho \propto \eta$, where $\eta$
is the viscosity of the electron fluid in the absence of disorder.

The best analogy is with the viscosity of liquid $^3$He and $^4$He
in the same  temperature range. The experimental fact that
$\rho\sim 1/T$ in this temperature interval (see
Fig.\ref{fig:fig2} and \textcite{mills99})  is not inconsistent with the conjecture, Eq.~\ref{voscosity},
that the viscosity of liquid He in the same ``semiquantum''
regime, obeys $\eta\sim 1/T$.  (Measured values \cite{AndreevSQL1}
of the viscosity in bulk $^4$He are consistent with $\eta \sim
1/T$, but as far as we know, no comparable data exists  in bulk
$^3$He, nor in $^3$He or $^4$He films.)

\subsection{Interpretation of experiments based on percolation}

Percolation is a classical concept whose relevance
to quantum
 systems requires a separation of spatial scales \cite{ShklovskiiEfros}.  Specifically,  in order that different (randomly distributed) regions of the sample can be treated as sufficiently macroscopic 
 to be characterized by a local value of the conductivity, the correlation length of the scattering potential
must be large compared to the relevant microscopic lengths needed to define a local electronic phase.  At the very least, percolation is a useful concept only if the disorder potential varies on distances large compared to the distance between electrons. In particular, since the quantum dephasing length, $L_\phi$ diverges as $T\to 0$, percolation  is never  justified at very low $T$ because it neglects the localization,
interference effects, etc. in the "metallic" state.  So percolation at best provides a valid description at intermediate temperatures. However, if the disorder has sufficiently long-range correlations, percolation effects can readily
mimic  the finite $T$
 appearance of a metal-insulator transition.

Such a theory can be developed
in a form which is applicable to samples with small $r_{\text s}$
in the presence of a smooth scattering potential. (This situation can be realized, for example, in 2D samples where charged impurities are separated
from the 2D electron liquid by a wide spacer.) In this case, to describe the transition one has to take into account non-linear screening \cite{Efros,Fogler}.

\textcite{Sarma1,Sarma2} applied the 
same sort of theory, extrapolated to large $r_{\text s}$, to
explain experimental data for low density electron liquids in
 in the
metallic state near the point of the metal-insulator transition.
Again, despite the reported qualitative successes of this theory, we believe that there are significant problems with its straightforward application in the current context. For example,  as $B_{\|}$ increases the Fermi energy of weakly interacting electron liquid increases, ultimately by a factor of 2.  Generally, this would be expected to {\it increase} the faction of the sample that is in the metallic phase. Therefore such a theory would 
result in a resistance which is a decreasing function of increasing
 $B_{\|}$, with even the possibility that it could induce an insulator to metal transition.  Conversely,  the experimental data
presented in Fig.4  show  a field-driven metal to insulator transition.

Indeed,
the assumption of very long-range correlations of the disorder is problematic in itself.  Certainly, in Si-MOSFETs, it is generally accepted that the disorder has rather short-range correlations.  More generally, the distance to the remote dopants is rarely significantly longer than the typical distance between electrons.  Thus, the large separation of length scales assumed in a simple percolation analysis cannot be ubiquitous among the devices studied.

On the other hand, the metal-Wigner crystal transition (and associated intermediate micro-emulsion phases) in the presence of weak disorder yields a state which can naturally be described by percolation through the metallic fraction of the system. Even for short-range correlated disorder, the familiar disorder broadening of a first order transition can lead to ``Imry-Ma'' domains of large size (\textcite{Imry}; \textcite{Aizenman}; \textcite{Baker}).  Although, at the end of the day,  a percolation picture based, in this way, on disorder induced phase coexistence has many features in common with the single-particle percolation picture 
 \cite{Efros,Fogler,Sarma1,Sarma2}, above, it has the important differences that:  (i)  The large length scales that justify the percolation analysis are self-organized, rather than inherited directly from the disorder potential, and (ii)  The resulting pattern of insulating and conducting phases is a strong function of any parameter, especially $T$ and $B_{\|}$, which affects the thermodynamic balance between the two phases.

As we have already discussed,
this general type of theory is in  qualitative agreement with much of the experimental data. For example, due to the Pomeranchuk effect, the fraction of the Wigner crystal increases with increasing $B_{\|}$, which in turn can lead to a field-induced transition from a metal to an insulator.
Another advantage of consideration of this regime is that it 
offers a simple explanation of why highly conducting samples can still have such strongly $T$ and $H_{\|}$ dependent conductivities.
 Unfortunately, however, no quantitative theory of the transport in this regime  currently exists.

\subsection{Experiments on double-layers}

We now turn to the anomalies in the measured
 drag resistance $\rho_{\text {D}}$ \cite{Pillarisetty} in p-GaAs double layers
 with small resistances  and large $r_{\text s}$.

A straightforward application of the Fermi liquid based theory valid at $r_{\text s}\ll 1$
yields results in qualitative contradiction with experimental results of \textcite{Pillarisetty}. For example since in the framework of the Fermi liquid theory the electron scattering rate is inversely proportional to the Fermi energy, it natural to expect that the drag resistance decreases as $B_{\|}$ increases. This is in contradiction with the experiment.

It was shown by \textcite{Kamenev} that
interference corrections to $\rho_{\text {D}}$
are increasingly important at low $T$ and indeed do not vanish as $T\to 0$.
 This is in loose qualitative agreement with the fact that the experimental values of the drag resistance are much larger
  than the conventional theoretical values~\cite{Price,MacDonnald}, which vanish in proportion to $T^2$ as $T \to 0$.
  Secondly, according to \textcite{Kamenev},
$\rho_{\text {D}}\sim \rho^{3}$. This is also in qualitative agreement with experiment where both $\rho_{\text {D}}(B_{\|})$ and $\rho^{3}(B_{\|})$ increase with $B_{\|}$ and saturate at $B_{\|}>B^{*}$. On the other hand, as we have already discussed, since this theory
was derived in the limit $r_{\text s}\ll 1$ and $k_{\text F}l\gg 1$, it cannot answer the more basic question of   why $\rho(T, B_{\|})$ itself increases as a function of $T$ and $B_{\|}$, nor why there is a  maximum of $\rho(T)$.

A focused study based on an extrapolation of Fermi liquid based formulas  to the region where $r_{\text s}\gg 1$ has been carried out by \textcite{SarmaDrag} in order to address the drag resistance data of \textcite{Pillarisetty}.
 According to \textcite{SarmaDrag}, the large enhancement of the magnitude of the drag resistance arises from a combination of a number of separate (somewhat technical) features of the specific experimental system in question.  Moreover, these authors find that both the resistance
of individual layers, and the the drag magneto-resistance turn out to be
increasing functions of $B_{\|}$.   As we have mentioned 
our concern here is with the underlying assumptions in this approach:  Despite its many successes,
the formulas for the electron cross-section used by \textcite{SarmaDrag}
were derived under an assumption that the standard expressions for electron screening continue to apply when the screening length, $\lambda_{sc}$
  is much smaller than the inter-electron distance. Specifically, in this approach,
the change of the sign of the drag magneto-resistance which occurs on extrapolating the small $r_{\text s}$ results to large $r_{\text s}$ can be traced
to the fact that
$\lambda_{sc}\ll n^{-1/2}$, which is clearly unphysical.

An attempt to explain the experimental data \cite{Pillarisetty} based on the assumed existence of  microemulsion phases was carried out by \textcite{SpivakKivelsonDrag}. The main assumption 
is that a phase consisting of {\it mobile} bubbles of Wigner crystal embedded in the Fermi liquid is responsible for the large drag resistance. Since the bubbles have different electron densities than the surrounding Fermi liquid, they produce a large electric potential which is seen by  the electron liquid in the second layer, thus producing a large $\rho_D$. However, the existence of  mobile bubbles  at the relevant temperatures is an assumption that has not yet been tested, since no reliable theoretical estimates have yet been made of the characteristic size of the bubbles, nor the range of densities over which the bubbles are stable, and no direct experimental imaging of an electron microemulsion has yet been achieved. Once this assumption is accepted, all qualitative features of experiment \cite{Pillarisetty} can be explained as consequences of the Pomeranchuk effect.

\subsection{$n$-dependence of $B^{*}$ }

In discussing the $n$ dependence of $B^{*}$, it is again useful to
compare it to the analogous problem in $^3$He. The problem of the
density dependence of the saturation magnetic field and
enhancement of the spin susceptibility in the strongly correlated
Fermi liquid $^{3}$He near the crystallization transition has been
discussed by \textcite{kostang}. There, it was pointed out that
two different scenarios can be imagined with different
consequences for the evolution of the magnetic response:\\
(i) The
system is nearly ferromagnetic, which means that the linear spin
susceptibility is enhanced compared to its non-interacting value,
but the saturation field, $B^{*}$, is not suppressed. This would
mean that the system is close to a Stoner instability.\\
(ii) The
system is nearly solid. In this case, both the linear and all
non-linear susceptibilities are enhanced, and $B^{*}$ is
suppressed compared to the non-interacting values.

It seems to us
that the data in the large $r_{\text s}$ 2DEG are more generally
consistent with the second scenario.

 \section{Conclusion}
\label{conclusion}

We have summarized a large body of  experimental data and theoretical arguments
which strongly imply the existence  of qualitatively new (``non-Fermi liquid'') physics of the 2DEG in the limit of large  $r_{\text s}$ and weak disorder.
These phenomena include,  but are not limited to those associated with  a metal insulator transition.
We have also critically discussed some of the attempts to make contact between theory and experiment concerning these behaviors.

The primary purpose of this article is to bring into focus the problems of physics
which are unresolved, and to stimulate further study, especially experimental study of these problems.  Because of the absence of a well controlled theory of the metal insulator transition in a strongly correlated 2DEG, there has been  a large amount of controversy surrounding the interpretation of these experiments.  However, it is important to stress that there is no controversy concerning the experimental facts, especially given that similar phenomena are seen by multiple groups in a variety of very different types of devices.

We conclude our discussion with a brief list of some of the properties of the 2DEG which could readily be measured, but which have not (to the best of our knowledge) been seriously studied to date.
More generally, the widely celebrated progress that is ongoing in the fabrication and broader distribution of increasingly high mobility Si and (especially) GaAs based devices
 will open the possibility of probing the intrinsic properties of clean 2DEGs at still lower densities and temperatures, where presumably all the previously observed behaviors will have still larger amplitudes.
 Moreover, as high mobility devices made with other semiconductors become available, further tests of the universality of the phenomena, as well as interesting particulars associated with differences in band-structure and the like, could add  to our knowledge of this subject.

 (i)
 There is surprisingly little data concerning the properties of strongly correlated  electron liquids in the semi-quantum regime $E_{\text F}< T<\sqrt{r_{\text s}}E_{\text F}\sim \sqrt{E_FV}$, where it is non-degenerate, but still highly quantum mechanical, and in the classical regime, $\sqrt{r_{\text s}}E_{\text F} < T < r_{\text s} E_{\text F} \sim V$, where it is still highly correlated but quantum effects are negligible.
  For example,
 experimental data
on the magneto-resistance in a parallel magnetic field at $T>E_F$
 have not been reported for any of the materials in question.
 (It is also worth noting that systematic studies of the viscosity of  $^3$He, especially $^3$He films, in the semi--quantum regime do not, to the best of our knowledge, exist and would be useful for comparison.)

 (ii) As the resistance is the most directly measured property of the
2DEG, most empirical information about its character comes from
transport measurements.  However, thermodynamic properties such as
compressibility can be measured on 2DEGs \cite{EisensteinCompress}.
There have been some recent measurements of compressibility
showing a drastic difference between the metallic and insulating
phases (\textcite{Jiang}; \textcite{Savchenko}; \textcite{yacoby}). It would be desirable,
however, to perform more studies in samples with even higher
mobility and $r_{\text s}$ than those used by \textcite{Jiang}, \textcite{Savchenko}, and \textcite{yacoby} to
elucidate the difference between high $r_{\text s}$ samples with strong
metallic transport and the 2DEG's with lower
$r_{\text s}$ \cite{EisensteinCompress}.

(iii) Thermoelectric and Nernst effects can reveal the strongly correlated nature
 of an electron system. However, we are aware of only a very limited  number of papers (\textcite{ShayeganSeebeck}; \textcite{Shayegan}; \textcite{FletcherSeebeck}) reporting thermoelectric measurements in strongly correlated 2DEGs, while the Nernst effect has never been measured in devices with large $r_{\text s}$.

(iv) Since drag effects are uniquely sensitive to charge density fluctuations in the 2DEG, further experiments  in double layer systems, especially in the presence of perpendicular, and (even more importantly) parallel magnetic fields, could be very illuminating. For example the drag resistance in the semi-quantum regime $T>T_{\text F}$ has never been measured.

(v) While isolated experiments exist on the quantum Hall effect in devices with moderately large $r_{\text s}$, systematic studies of the evolution from the quantum Hall states to the Wigner crystal state as a function of increasing $r_{\text s}$ do not exist.  Moreover, as discussed above, there is much to be learned about the phases and phase transitions at $B_\perp=0$ by following the evolution of the various quantum Hall phases to small $B_\perp$ in clean devices with large $r_{\text s}$.

(vi) It would be very desirable to extend the few existing experiments \cite{eisenstein,StormerZhuSpin} on low density, large $r_{\text s}$ electron gases in ultra-clean n-GaAs devices, and in particular to measure effects of the magnetic field.

(vii)  One very promising avenue for obtaining more local information about the nature of the 2DEG at large $r_{\text s}$, more or less free of the complications due to quenched disorder, is to study their properties in mesoscopic devices such as ``point contacts''. Up until now, however, they have primarily been studied only in three regimes: the regime where electron interactions are not important \cite{Wees}, the Coulomb blockade regime, and the Kondo regime \cite{GoldhaberGordon,RikhGlazman} in which there is a single localized electronic state through which tunneling between the two metallic reservoirs occurs. However, there is another interesting regime when there is a relatively large ``depletion region''  between two metallic reservoirs with high electron densities.  In the depletion region, there is a  strongly correlated electron liquid with low electron density whose density can be varied by changing a gate voltage. In this case as a function of gate voltage, aspects of the electronic micro-emulsion phases can be directly probed on a mesoscopic scale. In this context we would like to mention that, as discussed by \textcite{SpivKivAnn}, the Pomeranchuk effect can produce significant $T$ and $B_{\|}$  dependences to the resistivity through the depletion region. It is interesting that it can produce dependences of $\rho(T, B_{\|})$ which can mimic some of the behavior traditionally associated with the Kondo effect.

(viii)  There are various new experimental methods
(\textcite{ashoori98}; \textcite{yacoby_new}; \textcite{goldhaber}) being developed which hold the
promise of providing spatially resolved images of the evolving
physics of the 2DEG at large $r_{\text s}$.  Needless to say, such data
could revolutionize our understanding of the strong correlation
effects in the 2DEG.

(ix)  Finally, although it may be a while until sufficiently well ordered materials are available for these purposes, it is likely that  advances in the study of double-layer graphene and related materials may open unprecedented opportunities to study the properties of the 2DEG at large $r_{\text s}$.
 Due its Dirac spectrum, $r_{\text s}$ does not depend on $n$  in a single layer of graphene,  and more generally correlation effects are quite different than in problems with a quadratic dispersion.  However, bilayers of graphene have a quadratic dispersion, so $r_{\text s}\to \infty$ as $n \to 0$.

\begin{acknowledgments}
We thank E. Abrahams, G.~S. Boebinger, S. Chakravarty, S. Das~Sarma, A.~M. Finkel'stein, J. Folk, H.-W. Jiang, D.~E. Khmelnitskii, P.~A. Lee, L. Levitov, S. Maslov, M.~P. Sarachik, A.~A. Shashkin, B.~I. Shklovskii, D. Shahar, D.~C. Tsui, V.~M. Pudalov, and S.~A. Vitkalov for useful discussions.
BS was supported by  NSF grant DMR-0704151.  SVK was supported by DOE Grant DE-FG02-84ER45153 and by BSF grant \# 2006375. SAK was supported by DOE Grant DE-AC02-76SF00515. XPAG was supported by CWRU startup fund.
\end{acknowledgments}



\end{document}